\documentclass[journal]{IEEEtran}

\usepackage{amsmath,amssymb,amsfonts,mathtools,bm,amsthm}
\usepackage{graphicx}
\usepackage{cite}
\usepackage{booktabs}
\usepackage{multirow}
\usepackage{xcolor}
\usepackage{epstopdf}
\usepackage{array}
\usepackage{algorithm}
\usepackage{placeins}
\usepackage{tikz}
\usepackage{pgfplots}
\usepackage{algpseudocode}

\providecommand{\jj}{\mathrm{j}}
\DeclareMathOperator{\diag}{diag}

\DeclareMathOperator{\vecop}{vec}
\DeclareMathOperator{\tr}{tr}

\DeclareMathOperator*{\argmin}{arg\,min}
\DeclareMathOperator*{\argmax}{arg\,max}

\newtheorem{proposition}{Proposition}
\newtheorem{lemma}{Lemma}

\begin{document}

\title{Grouped Sparse Bayesian Estimation for Near-Field Massive MIMO Channels}

\author{Xingchen Fang, Lan Tang,~\IEEEmembership{Member, IEEE}, Wei Xu,~\IEEEmembership{Fellow, IEEE}%
\thanks{This work was supported by National Natural Science Foundation of China (U25A20393) and Jiangsu Provincial `Qing Lan Project'. (Corresponding author: Lan Tang.)}%
\thanks{Xingchen Fang, Lan Tang are with the School of Electronic Science and Engineering, Nanjing University, Nanjing, China (e-mail: tanglan@nju.edu.cn).}%
\thanks{Wei Xu is with the National Mobile Communications Research Laboratory, Southeast University, Nanjing, China.}}

\maketitle

\begin{abstract}
This paper studies near-field massive MIMO channel estimation based on an electromagnetic channel model with limited pilots. Starting from Maxwell's equations, we derive the homogeneous background response and the scattering response in the spatial and wavenumber domains. It is found that a compact scattering cluster produces a localized receive--transmit transverse-wavenumber block across finite apertures, thereby yields a block-sparse NLOS channel coefficient matrix in the wavenumber domain. Considering the spatial non-stationarity of large-scale antenna arrays, we introduce visibility factors at the transmit and receive sides of the channel model. To solve the non-stationary sparse channel estimation problem in the wavenumber domain, we propose a visibility-aware block-sparse Bayesian channel estimation algorithm. Simulation results demonstrate that the proposed method achieves lower channel-estimation NMSE than benchmark schemes.
\end{abstract}


\section{Introduction}

Massive multiple-input multiple-output (MIMO) has been one of the key physical-layer technologies for modern wireless systems, and its extension toward extremely large-scale MIMO (XL-MIMO), holographic MIMO, and high-frequency arrays is expected to provide substantially higher spatial resolution and multiplexing capability \cite{Marzetta2010,Larsson2014,Rusek2013,Hu2018,Wang2024Tutorial}. As the aperture becomes electrically large and the operating wavelength becomes small, many receivers and scatterers may lie in the radiating near-field region. In this regime, the classical far-field plane-wave approximation is no longer sufficient: the channel depends jointly on angle and distance, the wavefront is spherical rather than planar, and the received field may become spatially non-stationary across different parts of the aperture \cite{Bjornson2020,LuZeng2022,CuiDai2022,An2023Survey,Liu2023Tutorial}. These properties make near-field channel estimation fundamentally different from conventional far-field sparse estimation. Recent surveys review the resulting channel models and estimation challenges \cite{Lu2024Tutorial,Liu2025Survey}.

Most existing communication-oriented near-field channel estimation methods are built on geometric or parametric channel representations, such as spherical-wave steering vectors, polar-domain dictionaries, distance--angle grids, or hybrid near-/far-field codebooks \cite{CuiDai2022,WeiDai2022,CuiDai2023Wideband,Zhang2023NF,DaiMixed2023}. These models have enabled efficient compressed-sensing and stagewise estimation algorithms, including the two-stage mixed-LOS/NLOS method of Lu and Dai \cite{DaiMixed2023}. Spatially non-stationary channel estimation also accounts for the different propagation paths observed by different subarrays \cite{ChenDai2024Nonstationary}. However, they usually take the array response or the channel matrix itself as the primitive object. As a result, source excitation, electromagnetic field generation, scattering, polarization, receive-port formation, and local visibility are absorbed into a prescribed input--output parametrization. This abstraction is useful algorithmically, but it obscures how the estimator is connected to the underlying electromagnetic propagation mechanism.

In parallel, electromagnetic-domain and wavenumber-domain models have attracted increasing attention for holographic and near-field MIMO systems \cite{Pizzo2020,Pizzo2022FPWS,Pizzo2022Nyquist,Demir2022,Gong2024JSAC,WanDai2024EIT}. Fourier plane-wave expansions and electromagnetic-domain channel models provide physically meaningful descriptions of wave propagation, spatial degrees of freedom, and capacity limits. Related electromagnetic information theory studies further show that Maxwell-consistent modeling is essential for understanding electrically large apertures and continuous surfaces \cite{Pizzo2022FPWS,Gong2024JSAC,WanDai2024EIT,Miller2019}. Nevertheless, many of these works are primarily modeling-, capacity-, or evaluation-oriented. They did not directly provide a pilot-aided sparse estimator for grouped mixed LOS/NLOS near-field channels. Therefore, there remains a gap between continuous electromagnetic forward modeling and a finite-dimensional estimator that can be implemented in practical channel training.

Sparse recovery methods estimate a small number of active coefficients from limited observations \cite{Tropp2010}. Sparse Bayesian learning (SBL) learns coefficient variances to identify the active components \cite{Wipf2004,Ji2008BCS}. Further structure can be exploited through common channel support across subcarriers \cite{Gao2015Common} or correlation among source vectors in multiple-measurement models \cite{ZhangRao2011}. These results motivate the use of structured sparse estimation for the wavenumber-domain channel.

This paper connects an antenna-level electromagnetic forward model and its grouped implementation to a finite-dimensional sparse channel estimator. The continuous propagation channel model is first derived from antenna currents and the inhomogeneous wave equation. The resulting finite-aperture response is then represented on a transverse-wavenumber grid and mapped to a grouped low-complexity observation model under the grouped-pilot configuration. Comparisons with the geometric two-stage method in \cite{DaiMixed2023} are presented under the same grouped observations, pilot matrix, and specified LOS/NLOS power setting.

The main contributions are summarized as follows.
\begin{itemize}
    \item We establish a near-field electromagnetic channel model that explicitly connects antenna excitation, wave propagation, scattering, visibility, and the received signal. The model describes both LOS and NLOS fields and supports transmit/receive antenna grouping for low-complexity estimation.
    \item We reveal the group-sparse structure of the electromagnetic channel in the wavenumber domain. The finite-aperture response of each compact scattering cluster is concentrated in a local receive--transmit wavenumber block and therefore leads to a block-sparse NLOS representation.
    \item We therefore develop a visibility-aware block-SBL algorithm for the grouped channel model. The algorithm alternately estimates the LOS/NLOS coefficients and the bidirectional Markov visibility on a conservative candidate support, followed by a restricted coefficient refit. A hybrid Cram\'er--Rao--Weiss--Weinstein bound (HCRWWB) with deterministic continuous parameters and random LOS/NLOS visibility is also derived as a theoretical performance lower bound for comparison.
    \item Simulation results show that the proposed method achieves lower NMSE than spatial near-field OMP and wavenumber block-OMP, and approaches the HCRWWB as the pilot length increases. The results also show the advantages of antenna grouping in achieving a performance--complexity tradeoff.
\end{itemize}

The rest of this paper is organized as follows. Section~II presents the massive MIMO channel model with transmit/receive antenna grouping. Section~III develops the electromagnetic channel model in the wavenumber domain. Section~IV presents the bidirectional visibility-aware block-sparse Bayesian estimator and analyzes its complexity. Section~V presents the numerical results, and Section~VI concludes the paper. Appendix~\ref{app:hcrwwb_derivation} provides the detailed derivation and evaluation of the HCRWWB.

\section{Massive MIMO Channel Model with Transmit/Receive Antenna Grouping}

To reduce the complexity of channel estimation in massive MIMO, this section adopts an antenna-grouping-based transmit/receive scheme, where adjacent transmit and receive antennas are represented by prescribed groups. Consequently, channel estimation between individual antenna pairs is reduced to that between antenna groups. The effect of grouping on estimation accuracy is evaluated in the simulation section.

\begin{figure}[!htbp]
\centering
\includegraphics[width=\columnwidth]{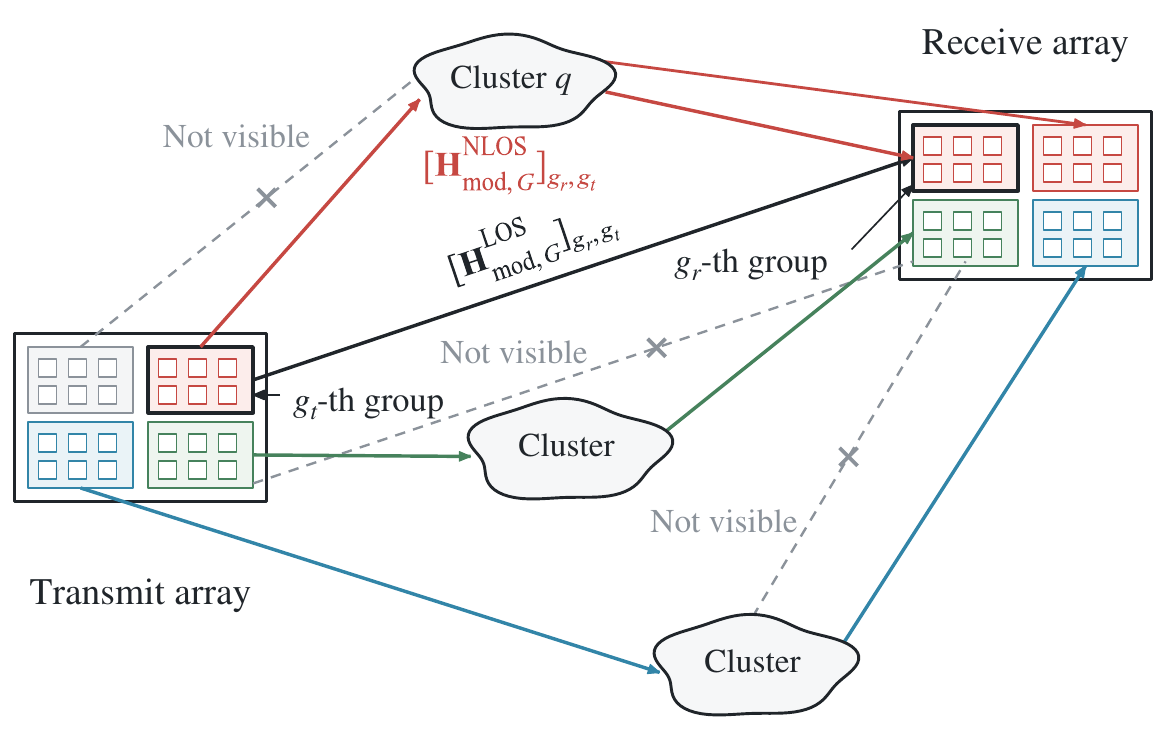}
\caption{Near-field massive MIMO system model. }
\label{fig:system_model}
\end{figure}

Consider a near-field massive MIMO link with \(N_t\) transmit antennas and \(N_r\) receive antennas. The transmit and receive arrays are partitioned into \(G_t\) and \(G_r\) groups, respectively. Let \(\mathcal U^t_{g_t}\) and \(\mathcal U^r_{g_r}\) denote the antenna-index sets for the \(g_t\)-th transmit antenna group and the \(g_r\)-th receive antenna group, respectively.
Define \(\mathbf r_{n_r},\mathbf r_{n_t}\in\mathbb R^3\) as the physical positions of receive antenna \(n_r\) and transmit antenna \(n_t\), respectively. The propagation environment consists of a direct LOS component and a small number of effective NLOS scattering clusters located at positions \(\mathbf c_q=[x_q,y_q,z_q]^T\in\mathbb R^3\), \(q=1,\ldots,Q_c\). For electrically large arrays, different subarrays may observe different propagation components because of local blockage, cluster visibility, and spherical-wave propagation \cite{WangJSAC2021}. This spatial non-stationarity motivates the introduction of visibility variables. In the grouped implementation, antennas in the same group are represented by the same group-center spatial response and visibility state. In Section~III, the continuous visibility variables \(\nu_{t,q}(\mathbf r_{n_t})\) and \(\nu_{r,q}(\mathbf r_{n_r})\) for the \(n_t\)-th transmit and \(n_r\)-th receive antennas with respect to the \(q\)-th scatterer are introduced first, based on which their group-level counterparts are further defined.

Assume that the training block contains \(P\) pilot symbol periods. The grouped pilot matrix is
\begin{equation}
\mathbf S=
\big[
\mathbf s[0],\mathbf s[1],\ldots,\mathbf s[P-1]
\big]
\in\mathbb C^{G_t\times P},
\label{eq:Sdef}
\end{equation}
where \(\mathbf s[p]=[s_1[p],\ldots,s_{G_t}[p]]^T\) is the grouped transmit pilot vector during the \(p\)-th symbol period. If transmit antenna \(n_t\in\mathcal U^t_{g_t}\), then its pilot is \(s_{n_t}[p]=\beta_{n_t,g_t}s_{g_t}[p]\), where \(\beta_{n_t,g_t}=|\mathcal U^t_{g_t}|^{-1/2}\), and it gives unit-norm equal excitation within each group. The equivalent baseband current in the \(g_t\)-th transmit group is

\begin{equation}
\tilde i_{g_t}(t)
=
\sum_{p=0}^{P-1}s_{g_t}[p]g(t-pT_s),
\label{eq:igt_t}
\end{equation}
where \(T_s\) is the symbol interval and \(g(t)\) is the pulse-shaping waveform.

\section{Near-Field Wavenumber-Domain Sparse Channel Modeling}

High-frequency near-field channels can exhibit significant spatial sparsity. Therefore, spatial-domain near-field multipath sparse models usually describe the channel by discretizing range--angle pairs or scatterer locations, and then fitting the received array response with a small number of geometric atoms \cite{CuiDai2022,WeiDai2022,CuiDai2023Wideband,DaiMixed2023}. This approach is intuitive for isolated point scatterers, but it becomes restrictive for electrically large apertures. Practical NLOS components are often generated by compact or extended scattering clusters rather than ideal point scatterers, rendering a purely point-based sparse representation less plausible and insufficient for accurately characterizing the scattering behavior of extended structures. More fundamentally, a prescribed spatial dictionary directly maps scatterer locations to channel vectors, yet it fails to explicitly incorporate source excitation, field propagation, contrast-induced scattering, and finite-aperture observation into channel model.

To address these limitations, we derive the wavenumber-domain channel model from the electromagnetic propagation process rather than imposing a spatial dictionary at the outset. Starting from Maxwell's equations under a grouped scalar-port approximation, the derivation yields the propagation relation for the inhomogeneous background.
After restricting the resulting continuous LOS/NLOS responses to the transmit and receive apertures, the transverse-wavenumber representation reveals a physically
motivated sparsity structure:
the dominant smooth-body contribution of a compact scattering cluster is concentrated in a local receive-transmit wavenumber block, up to finite-aperture and visibility-boundary leakage, rather than on an arbitrary set of spatial grid points. This underpins the proposed block-sparse wavenumber-domain estimator developed in Sec. IV.

%that the response of a compact scattering cluster is concentrated in a local receive--transmit wavenumber block. This structure underpins the proposed block-sparse wavenumber-domain estimator.

\subsection{Wavenumber-Domain Transmission Equation for a Homogeneous Medium}

In this subsection, we start from Maxwell's equations and derive the transmission equation in the wavenumber domain for a homogeneous medium. For transmit antenna \(n_t\) at \(\mathbf r_{n_t}\) and a receiving point \(\mathbf r\), Maxwell's equations in a homogeneous isotropic background medium are written as
\begin{align}
\nabla\times\mathbf E(\mathbf r,\mathbf r_{n_t},t)
&=
-\mu_b
\frac{\partial\mathbf H(\mathbf r,\mathbf r_{n_t},t)}{\partial t},
\label{eq:maxwell_faraday}
\\
\nabla\times\mathbf H(\mathbf r,\mathbf r_{n_t},t)
&=
\varepsilon_b
\frac{\partial\mathbf E(\mathbf r,\mathbf r_{n_t},t)}{\partial t}
+
\mathbf J(\mathbf r,\mathbf r_{n_t},t),
\label{eq:maxwell_ampere}
\\
\nabla\cdot\big(\varepsilon_b\mathbf E(\mathbf r,\mathbf r_{n_t},t)\big)
&=
\rho_e(\mathbf r,\mathbf r_{n_t},t),
\label{eq:maxwell_gauss_e}
\\
\nabla\cdot\big(\mu_b\mathbf H(\mathbf r,\mathbf r_{n_t},t)\big)
&=
0,
\label{eq:maxwell_gauss_h}
\end{align}
where \(\mathbf E(\mathbf r,\mathbf r_{n_t},t)\), \(\mathbf H(\mathbf r,\mathbf r_{n_t},t)\), \(\mathbf J(\mathbf r,\mathbf r_{n_t},t)\), and \(\rho_e(\mathbf r,\mathbf r_{n_t},t)\) denote
the vector electric field, magnetic field, impressed current density, and free charge density at a receiving point \(\mathbf r\) due to transmit antenna \(n_t\), respectively. The operator \(\nabla\) differentiates with respect
to \(\mathbf r\).
Here \(\varepsilon_b\) is the background permittivity and \(\mu_b\) is the background permeability.

Based on \eqref{eq:maxwell_faraday} and \eqref{eq:maxwell_ampere}, the electric-field double-curl equation is obtained as
\begin{equation}
\nabla\times\nabla\times\mathbf E(\mathbf r,\mathbf r_{n_t},t)
=
-\mu_b\varepsilon_b
\frac{\partial^2\mathbf E(\mathbf r,\mathbf r_{n_t},t)}{\partial t^2}
-
\mu_b
\frac{\partial\mathbf J(\mathbf r,\mathbf r_{n_t},t)}{\partial t}.
\label{eq:curl_derivation}
\end{equation}
In a source-free propagation region away from the transmit antenna, we have \(\nabla\cdot\mathbf E(\mathbf r,\mathbf r_{n_t},t)=0\). Applying the vector identity \(\nabla\times(\nabla\times\mathbf E)=\nabla(\nabla\cdot\mathbf E)-\nabla^2\mathbf E\) to \eqref{eq:curl_derivation}, the electric-field double-curl equation reduces to the vector wave equation.

Each receive antenna measures a projected scalar electric field. The corresponding scalar wave equation is
\begin{equation}
\nabla^2 e(\mathbf r,\mathbf r_{n_t},t)
-
\mu_b\varepsilon_b
\frac{\partial^2e(\mathbf r,\mathbf r_{n_t},t)}{\partial t^2}
=
\mu_b
\frac{\partial j_s(\mathbf r,\mathbf r_{n_t},t)}{\partial t},
\label{eq:scalar_wave_time}
\end{equation}
where \(e(\mathbf r,\mathbf r_{n_t},t)\) is the projected scalar electric-field component sensed at \(\mathbf r\) due to transmit antenna \(n_t\), \(j_s(\mathbf r,\mathbf r_{n_t},t)\) is the corresponding effective scalar source term, and \(\nabla^2=\partial^2/\partial x^2+\partial^2/\partial y^2+\partial^2/\partial z^2\) is the Laplacian operator. For a single-source derivation, set the source term in \eqref{eq:scalar_wave_time} as
\begin{equation}
j_s(\mathbf r,\mathbf r_{n_t},t)
=
J(\mathbf r,\mathbf r_{n_t})\tilde i_{n_t}(t)e^{\jj\omega_c t}
\end{equation}
where \(J(\mathbf r,\mathbf r_{n_t})=\delta^{(3)}(\mathbf r-\mathbf r_{n_t})\) is the space point-source and \(\omega_c\) is the carrier angular frequency.
Assume that \(\tilde i_{n_t}(t)\) has bandwidth much smaller than the carrier frequency and that the medium dispersion near \(\omega_c\) can be ignored. Under the slowly varying envelope approximation, the response field can be written as \(e(\mathbf r,\mathbf r_{n_t},t)\approx E(\mathbf r,\mathbf r_{n_t})\tilde i_{n_t}(t)e^{\jj\omega_c t}\). After differentiating \(j_s(\mathbf r,\mathbf r_{n_t},t)\) and \(e(\mathbf r,\mathbf r_{n_t},t)\) with respect to \(t\), and substituting them into \eqref{eq:scalar_wave_time}, the first- and second-order derivatives of \(\tilde i_{n_t}(t)\) can be ignored under the narrowband signal assumption. Removing the common factor \(\tilde i_{n_t}(t)e^{\jj\omega_c t}\) yields the spatial equation
\begin{equation}
\left(\nabla^2+k_c^2\right)E(\mathbf r,\mathbf r_{n_t})
=
\jj\omega_c\mu_b J(\mathbf r,\mathbf r_{n_t}),
\label{eq:homogeneous_spatial_helmholtz}
\end{equation}
where \(k_c=\omega_c\sqrt{\mu_b\varepsilon_b}\) is the carrier wavenumber.
For \(\mathbf r\neq \mathbf r_{n_t}\), the right-hand side of \eqref{eq:homogeneous_spatial_helmholtz} vanishes and the field satisfies the source-free Helmholtz equation.

The following derivation uses the spatial Fourier transform \(F(\mathbf k)=\int_{\mathbb R^3}f(\mathbf r)e^{-\jj\mathbf k^T\mathbf r}\,d\mathbf r\), with the convention \(\mathcal F\{\nabla^2 f\}=-\|\mathbf k\|^2F(\mathbf k)\). Applying the spatial Fourier transform to \eqref{eq:homogeneous_spatial_helmholtz} gives
\begin{equation}
\left(
k_c^2-\|\mathbf k\|^2
\right)E(\mathbf k,\mathbf r_{n_t})
=
\jj\omega_c\mu_b J(\mathbf k,\mathbf r_{n_t}),
\label{eq:scalar_wave_kw}
\end{equation}
where \(E(\mathbf k,\mathbf r_{n_t})\) and \(J(\mathbf k,\mathbf r_{n_t})\) are the corresponding spatial spectra. Therefore, the scalar background propagation relation is
\begin{equation}
E(\mathbf k,\mathbf r_{n_t})=G_b(\mathbf k)J(\mathbf k,\mathbf r_{n_t}),
\label{eq:background_transfer}
\end{equation}
where
\(G_b(\mathbf k)=\frac{\jj\omega_c\mu_b}{k_c^2-\|\mathbf k\|^2-\jj0}\);
the \(-\jj0\) term enforces the outgoing-wave condition. Note that \(\|\mathbf k\|=k_c\) is the pole of the Green function. Physically, it corresponds to propagating waves, i.e., waves that carry energy outward \cite{BornWolf1999}. The channel estimator
retains the propagating transverse wavenumber modes associated with this dispersion surface; evanescent and other near-source components outside
the retained set are included in the modeling residual. This
restriction is a fixed-frequency modal truncation rather
than a far-field approximation.

%transverse-wavenumber modes associated with this dispersion surface; evanescent and other near-source components outside the retained set are included in the modeling residual. This restriction does not invoke a far-field approximation.

\subsection{Wavenumber-Domain Transmission Equation for an Inhomogeneous Medium}
\label{subsec:inhomogeneous_decomposition}

A practical propagation environment, however, contains scattering inhomogeneities. Let
\begin{equation}
k^2(\mathbf r)=k_c^2+v(\mathbf r),
\label{eq:local_wavenumber}
\end{equation}
where \(v(\mathbf r)\) is the scattering contrast. The scalar inhomogeneous Helmholtz equation is
\begin{equation}
\left(\nabla^2+k_c^2\right)E(\mathbf r,\mathbf r_{n_t})
=
\jj\omega_c\mu_b J(\mathbf r,\mathbf r_{n_t})
-
v(\mathbf r)E(\mathbf r,\mathbf r_{n_t})
\label{eq:inhomogeneous_spatial}
\end{equation}
where \(v(\mathbf r)E(\mathbf r,\mathbf r_{n_t})\) is the contrast-induced scattering source term. This equation shows that the field in the inhomogeneous medium is equivalent to the field in a homogeneous medium excited by the impressed current density and the equivalent contrast sources arising from the scatterers \cite{Chew1995}. Define
\begin{equation}
V(\mathbf k)=\int_{\mathbb R^3}v(\mathbf r)e^{-\jj\mathbf k^T\mathbf r}\,d\mathbf r.
\label{eq:contrast_fourier_transform}
\end{equation}
Taking the spatial Fourier transform of \eqref{eq:inhomogeneous_spatial} gives
\begin{align}
E(\mathbf k,\mathbf r_{n_t})
&=
G_b(\mathbf k)J(\mathbf k,\mathbf r_{n_t})
\notag\\[-0.2em]
&\quad-
\frac{G_b(\mathbf k)}{\jj\omega_c\mu_b(2\pi)^3}
\int_{\mathbb R^3}
V(\mathbf k-\mathbf k_s)E_{\rm inc}(\mathbf k_s,\mathbf r_{n_t})\,d\mathbf k_s.
\label{eq:inhomogeneous_kw_convolution}
\end{align}
The first term in \eqref{eq:inhomogeneous_kw_convolution} is the LOS component, and the integral term is the NLOS component caused by the scatterers.

Under the single-scattering approximation, the field incident on the compact clusters can be approximated by the homogeneous incident field
\begin{equation}
E_{\rm inc}(\mathbf k_s,\mathbf r_{n_t})
\approx
G_b(\mathbf k_s)J(\mathbf k_s,\mathbf r_{n_t}).
\label{eq:incident_field_spectrum}
\end{equation}
The contrast \(v(\mathbf r)\) is associated with the spatial scattering clusters. For compact clusters, its spectrum is approximated by
\begin{equation}
V(\mathbf k-\mathbf k_s)
\approx
\sum_{q=1}^{Q_c}
\alpha_q
e^{-\jj(\mathbf k-\mathbf k_s)^T\mathbf c_q},
\label{eq:compact_contrast_spectrum}
\end{equation}
where \(\mathbf c_q\) is the center of the \(q\)-th cluster and \(\alpha_q\) is its effective scattering coefficient. Substituting \eqref{eq:compact_contrast_spectrum} and \eqref{eq:incident_field_spectrum} into \eqref{eq:inhomogeneous_kw_convolution} and applying the inverse spatial Fourier transform, the continuous NLOS response is given by
\begin{equation}
\begin{aligned}
&E^{\rm NLOS}(\mathbf r,\mathbf r_{n_t})
=-\sum_{q=1}^{Q_c}\!\alpha_q\nu_{r,q}(\mathbf r)\nu_{t,q}(\mathbf r_{n_t})
\biggl[(2\pi)^{-3}\!\int_{\mathbb R^3}G_b(\mathbf k)\\[-0.2em]
&{}\cdot e^{\jj\mathbf k^T(\mathbf r-\mathbf c_q)}d\mathbf k\biggr]
\biggl[(2\pi)^{-3}\int_{\mathbb R^3}
G_b(\mathbf k_s)e^{\jj\mathbf k_s^T(\mathbf c_q-\mathbf r_{n_t})}d\mathbf k_s\biggr]
+e_{\rm N}.
\end{aligned}
\label{eq:Enlos_spatial_response}
\end{equation}
The variables \(\nu_{t,q}(\mathbf r_{n_t})\) and \(\nu_{r,q}(\mathbf r)\) are introduced here to account for the fact that, due to spatial non-stationarity, the cluster sets visible to different transmit--receive pairs may not be the same.
Similarly, the spatial LOS response is
\begin{align}
E^{\rm LOS}(\mathbf r,\mathbf r_{n_t})
&=
\alpha_L
\nu_L(\mathbf r,\mathbf r_{n_t})
(2\pi)^{-3}
\notag\\[-0.2em]
&{}\cdot
\int_{\mathbb R^3}G_b(\mathbf k)
 e^{\jj\mathbf k^T(\mathbf r-\mathbf r_{n_t})}\,d\mathbf k
 +e_{\rm L}.
\label{eq:Elos_spatial_response}
\end{align}
Here \(\nu_L(\mathbf r,\mathbf r_{n_t})\) is the continuous LOS visibility, and \(\alpha_L\) is the common effective LOS coefficient. The terms \(e_{\rm N}\) and \(e_{\rm L}\) denote field-domain modeling residuals associated with the reduced LOS/NLOS representations.

Based on \eqref{eq:Enlos_spatial_response} and \eqref{eq:Elos_spatial_response}, compressed sensing with a spatial-domain dictionary can also be implemented by discretizing candidate scatterer locations and using the sampled LOS and NLOS Green responses as dictionary atoms. However, this requires a dense three-dimensional location grid, and the resulting near-field dictionary is large, highly coherent, and sensitive to off-grid mismatch, especially for electrically large apertures and compact scattering clusters. The following subsection instead derives the finite-aperture wavenumber representation.

\subsection{Block Sparsity in the Wavenumber Domain for Finite Arrays}
\label{subsec:compact_cluster_response}

Section~III-B gives the continuous source-to-destination response. We now restrict \eqref{eq:Enlos_spatial_response} to the finite transmit and receive apertures. Let \(\mathcal A_t\) and \(\mathcal A_r\) denote the two-dimensional transmit and receive aperture regions, respectively. The vectors \(\boldsymbol\rho_t\in\mathcal A_t\) and \(\boldsymbol\rho_r\in\mathcal A_r\) are their spatial coordinates, and \(\mathbf r_t(\boldsymbol\rho_t)\) and \(\mathbf r_r(\boldsymbol\rho_r)\) are the corresponding three-dimensional positions.

At the fixed carrier frequency, the propagating background modes satisfy \(\|\mathbf k\|=k_c\). Hence, a 3D wavevector can be parameterized by its transverse component \(\boldsymbol\kappa=[k_x,k_y]^T\) as \(\mathbf k^{\pm}(\boldsymbol\kappa)=[\boldsymbol\kappa^T,\pm\sqrt{k_c^2-\|\boldsymbol\kappa\|^2}]^T\), where the sign is selected according to the propagation direction. The spherical near-field response is retained through the superposition of transverse modes. Components outside the retained propagating region are included in the modeling residual.

Define the aperture-restricted continuous-domain visibility indicators by
\[
\nu_{r,q}^{A}(\boldsymbol\rho_r)
=\nu_{r,q}(\mathbf r_r(\boldsymbol\rho_r)),
\qquad
\nu_{t,q}^{A}(\boldsymbol\rho_t)
=\nu_{t,q}(\mathbf r_t(\boldsymbol\rho_t)).
\]
Here, ``continuous-domain'' refers to the aperture coordinates before antenna sampling. Substituting \(\mathbf r=\mathbf r_r(\boldsymbol\rho_r)\) and \(\mathbf r_{n_t}=\mathbf r_t(\boldsymbol\rho_t)\) into \eqref{eq:Enlos_spatial_response}, the finite-continuous aperture-domain NLOS field is expressed as
\begin{align}
E_{\mathcal A}^{\rm NLOS}(\boldsymbol\rho_r,\boldsymbol\rho_t)
&=
-\sum_{q=1}^{Q_c}\alpha_q
\nu_{r,q}^{A}(\boldsymbol\rho_r)
\nu_{t,q}^{A}(\boldsymbol\rho_t)
\notag\\[-0.2em]
&\quad\cdot
\chi_{r,q}(\boldsymbol\rho_r)
\chi_{t,q}(\boldsymbol\rho_t)
+e_{\rm N},
\label{eq:Enlos_aperture_continuous}
\end{align}
where
\begin{align}
\chi_{r,q}(\boldsymbol\rho_r)
&=
(2\pi)^{-3}
\int_{\mathbb R^3}G_b(\mathbf k_r)
 e^{\jj\mathbf k_r^T(\mathbf r_r(\boldsymbol\rho_r)-\mathbf c_q)}
 d\mathbf k_r
\notag\\[-0.2em]
&=-\frac{\jj\omega_c\mu_b}{4\pi}
\frac{e^{-\jj k_c\|\mathbf r_r(\boldsymbol\rho_r)-\mathbf c_q\|_2}}
{\|\mathbf r_r(\boldsymbol\rho_r)-\mathbf c_q\|_2},
\notag\\[-0.2em]
\chi_{t,q}(\boldsymbol\rho_t)
&=
(2\pi)^{-3}
\int_{\mathbb R^3}G_b(\mathbf k_t)
 e^{\jj\mathbf k_t^T(\mathbf c_q-\mathbf r_t(\boldsymbol\rho_t))}
 d\mathbf k_t
\notag\\[-0.2em]
&=-\frac{\jj\omega_c\mu_b}{4\pi}
\frac{e^{-\jj k_c\|\mathbf r_t(\boldsymbol\rho_t)-\mathbf c_q\|_2}}
{\|\mathbf r_t(\boldsymbol\rho_t)-\mathbf c_q\|_2},
\label{eq:chi_propagation_factors}
\end{align}
Similarly, the finite-continuous aperture-domain LOS field is expressed as
\begin{align}
E_{\mathcal A}^{\rm LOS}(\boldsymbol\rho_r,\boldsymbol\rho_t)
&=-\alpha_L\nu_L^A(\boldsymbol\rho_r,\boldsymbol\rho_t)
\frac{\jj\omega_c\mu_b}{4\pi}\notag\\[-0.2em]
&\quad\cdot\frac{e^{-\jj k_c\|\mathbf r_r(\boldsymbol\rho_r)-\mathbf r_t(\boldsymbol\rho_t)\|_2}}
{\|\mathbf r_r(\boldsymbol\rho_r)-\mathbf r_t(\boldsymbol\rho_t)\|_2}+e_{\rm L},
\label{eq:Elos_aperture_continuous}
\end{align}
where \(\nu_L^A(\boldsymbol\rho_r,\boldsymbol\rho_t)=\nu_L(\mathbf r_r(\boldsymbol\rho_r),\mathbf r_t(\boldsymbol\rho_t))\).
To reveal the wavenumber-domain block sparsity of the propagation response in \eqref{eq:Enlos_aperture_continuous}, take the transverse Fourier transforms of \(\chi_{r,q}(\boldsymbol\rho_r)\) and \(\chi_{t,q}(\boldsymbol\rho_t)\):
\begin{align}
\chi_{r,q}(\boldsymbol\kappa_r)
&=
\int_{\mathcal A_r}
w_r(\boldsymbol\rho_r)\chi_{r,q}(\boldsymbol\rho_r)
 e^{-\jj\boldsymbol\kappa_r^T\boldsymbol\rho_r}
 d\boldsymbol\rho_r,
\notag\\[-0.2em]
\chi_{t,q}(\boldsymbol\kappa_t)
&=
\int_{\mathcal A_t}
w_t(\boldsymbol\rho_t)\chi_{t,q}(\boldsymbol\rho_t)
 e^{\jj\boldsymbol\kappa_t^T\boldsymbol\rho_t}
 d\boldsymbol\rho_t,
\label{eq:chi_aperture_integrals}
\end{align}
where \(w_r(\boldsymbol\rho_r)\) and \(w_t(\boldsymbol\rho_t)\) are fixed, infinitely differentiable window functions with compact support inside the receive and transmit apertures, respectively. They suppress the wavenumber-domain leakage induced by finite-aperture truncation and thus yield a cleaner block-sparse representation in the wavenumber domain.

For \(u\in\{r,t\}\), assume \(R_{u,q}(\boldsymbol\rho_u)=\|\mathbf r_u(\boldsymbol\rho_u)-\mathbf c_q\|_2\ge d_0>0\). Substituting the spatial Green functions in \eqref{eq:chi_propagation_factors} into \eqref{eq:chi_aperture_integrals} and combining the propagation and Fourier phases yield
\begin{align}
\chi_{u,q}(\boldsymbol\kappa_u)
&=-\frac{\jj\omega_c\mu_b}{4\pi}
\int_{\mathcal A_u}\frac{w_u(\boldsymbol\rho_u)}{R_{u,q}(\boldsymbol\rho_u)}
e^{\jj k_c\psi_{u,q}(\boldsymbol\rho_u)}d\boldsymbol\rho_u
\notag\\[-0.2em]
&=\int_{\mathcal A_u}a_{u,q}(\boldsymbol\rho_u)
e^{\jj k_c\psi_{u,q}(\boldsymbol\rho_u)}
d\boldsymbol\rho_u,
\label{eq:oscillatory_aperture_form}
\end{align}
where \(a_{u,q}(\boldsymbol\rho_u)=-\frac{\jj\omega_c\mu_b}{4\pi}\frac{w_u(\boldsymbol\rho_u)}{R_{u,q}(\boldsymbol\rho_u)}\), \(\psi_{r,q}=-R_{r,q}-\boldsymbol\kappa_r^T\boldsymbol\rho_r/k_c\), and \(\psi_{t,q}=-R_{t,q}+\boldsymbol\kappa_t^T\boldsymbol\rho_t/k_c\).

The oscillatory integrals in \eqref{eq:oscillatory_aperture_form} have the following property, which locates the dominant wavenumber components.

\begin{lemma}[Nonstationary phase {\cite[Sec.~7.7]{Hormander2003}}]
\label{lem:oscillatory_localization}
If \(\|\nabla_{\boldsymbol\rho_u}\psi_{u,q}\|_2\ge\delta>0\) on the amplitude support, the asymptotic estimate of the oscillatory integral obtained by \(N+1\) integrations by parts satisfies
\begin{equation}
|\chi_{u,q}(\boldsymbol\kappa_u)|
\le C_{u,q,N,\delta}k_c^{-N},
\qquad k_c\to\infty,
\label{eq:oscillatory_nonstationary_decay}
\end{equation}
where \(N\) is any positive integer and \(C_{u,q,N,\delta}\) is independent of \(k_c\) for fixed geometry and aperture weights in the propagating region. Hence, the dominant high-frequency contributions come from neighborhoods of stationary points satisfying \(\nabla_{\boldsymbol\rho_u}\psi_{u,q}=\mathbf0\).
\end{lemma}

Without loss of generality, we set \(\mathbf r_u(\boldsymbol\rho_u)[1:2]=\boldsymbol\rho_u\). Then, based on Lemma~\ref{lem:oscillatory_localization}, we have the following proposition.

\begin{proposition}
\label{prop:finite_aperture_localization_main}
\textbf{Finite-aperture wavenumber-domain block sparsity.}
For the smoothly weighted response of cluster \(q\), the stationary conditions in Lemma~\ref{lem:oscillatory_localization} give the following receive and transmit wavenumber regions:
\begin{align}
\mathcal K_{r,q}
&=
\left\{
 -k_c
\frac{\boldsymbol\rho_r-\tilde{\mathbf c}_q}
{\|\mathbf r_r(\boldsymbol\rho_r)-\mathbf c_q\|_2}
:\boldsymbol\rho_r\in\mathcal A_r
\right\},
\label{eq:Krq_main}
\\[-0.2em]
\mathcal K_{t,q}
&=
\left\{
 +k_c
\frac{\boldsymbol\rho_t-\tilde{\mathbf c}_q}
{\|\mathbf r_t(\boldsymbol\rho_t)-\mathbf c_q\|_2}
:\boldsymbol\rho_t\in\mathcal A_t
\right\},
\label{eq:Ktq_main}
\end{align}
where \(\tilde{\mathbf c}_q=\mathbf c_q[1:2]\).
Lemma~\ref{lem:oscillatory_localization} shows that \(\chi_{u,q}(\boldsymbol\kappa_u)\) decays rapidly at high frequencies when \(\boldsymbol\kappa_u/k_c\) stays a fixed positive distance from \(\mathcal K_{u,q}/k_c\), \(u\in\{r,t\}\). Consequently, the dominant responses of the clusters in the receive-transmit transverse-wavenumber domain are concentrated near the union
\begin{equation}
\mathcal B_{\rm sp}
=
\bigcup_{q=1}^{Q_c}
\mathcal K_{r,q}\times\mathcal K_{t,q}.
\label{eq:Bsp_main}
\end{equation}
\end{proposition}

We now discretize the finite-continuous aperture-domain field to obtain its finite discrete representation. Let \(\{\boldsymbol\kappa_{r,i}\}_{i=1}^{Q_r}\) and \(\{\boldsymbol\kappa_{t,j}\}_{j=1}^{Q_t}\) denote the receive and transmit transverse-wavenumber grid points, respectively, and define \(\mathcal I_{r,q}=\{i:\boldsymbol\kappa_{r,i}\in\mathcal N(\mathcal K_{r,q})\}\) and \(\mathcal I_{t,q}=\{j:\boldsymbol\kappa_{t,j}\in\mathcal N(\mathcal K_{t,q})\}\), where \(\mathcal N(\mathcal K)\) denotes the retained grid points in and around \(\mathcal K\), and \(i\) and \(j\) index wavenumber atoms. The dominant wavenumber responses are represented by
\begin{align}
\chi_{r,q}(\boldsymbol\kappa_r)
&\approx
\sum_{i\in\mathcal I_{r,q}}
 b_{r,q,i}\delta(\boldsymbol\kappa_r-\boldsymbol\kappa_{r,i}),
\notag\\[-0.2em]
\chi_{t,q}(\boldsymbol\kappa_t)
&\approx
\sum_{j\in\mathcal I_{t,q}}
 b_{t,q,j}\delta(\boldsymbol\kappa_t-\boldsymbol\kappa_{t,j}).
\label{eq:chi_sparse_wavenumber}
\end{align}

The corresponding inverse transverse transforms are written as
\begin{align}
\chi_{r,q}(\boldsymbol\rho_r)
&=
\sum_{i\in\mathcal I_{r,q}}
 b_{r,q,i}e^{\jj\boldsymbol\kappa_{r,i}^T\boldsymbol\rho_r}
+e_{r,q}(\boldsymbol\rho_r),
\notag\\[-0.2em]
\chi_{t,q}(\boldsymbol\rho_t)
&=
\sum_{j\in\mathcal I_{t,q}}
 b_{t,q,j}e^{-\jj\boldsymbol\kappa_{t,j}^T\boldsymbol\rho_t}
+e_{t,q}(\boldsymbol\rho_t),
\label{eq:chi_aperture_synthesis}
\end{align}
where \(e_{u,q}(\boldsymbol\rho_u)\) includes the weighting difference \((1-w_u(\boldsymbol\rho_u))\chi_{u,q}(\boldsymbol\rho_u)\) and the wavenumber-grid approximation error, \(u\in\{r,t\}\).

The continuous visibility functions are sampled at the physical antenna positions to get \(\nu_{r,q}^{A}(\boldsymbol\rho_{r,n_r})\) and \(\nu_{t,q}^{A}(\boldsymbol\rho_{t,n_t})\), where \(\mathbf r_u(\boldsymbol\rho_{u,n_u})=\mathbf r_{n_u}\), \(u\in\{r,t\}\). Substituting \eqref{eq:chi_aperture_synthesis} into \eqref{eq:Enlos_aperture_continuous} and merging the cluster-wise coefficients gives the finite discrete NLOS field
\begin{align}
&E_{\mathcal A}^{\rm NLOS}(\boldsymbol\rho_{r,n_r},\boldsymbol\rho_{t,n_t})
=\sum_{q=1}^{Q_c}
\nu_{r,q}^{A}(\boldsymbol\rho_{r,n_r})
\nu_{t,q}^{A}(\boldsymbol\rho_{t,n_t})\notag\\[-0.2em]
&\cdot\sum_{i\in\mathcal I_{r,q}}\sum_{j\in\mathcal I_{t,q}}
x_{q,i,j}e^{\jj\boldsymbol\kappa_{r,i}^{T}\boldsymbol\rho_{r,n_r}
-\jj\boldsymbol\kappa_{t,j}^{T}\boldsymbol\rho_{t,n_t}}
+ [\boldsymbol\varepsilon_N]_{n_r,n_t}.
\label{eq:discrete_Enlos_synthesis}
\end{align}
where \([\boldsymbol\varepsilon_N]_{n_r,n_t}\) includes the residual terms from \eqref{eq:Enlos_aperture_continuous} and \eqref{eq:chi_aperture_synthesis}.
For each cluster, let \(\mathbf X_{N,q}\in\mathbb C^{Q_r\times Q_t}\) collect \(x_{q,i,j}\), and let \(\mathbf X_N=\sum_q\mathbf X_{N,q}\). Its support satisfies
\begin{equation}
\operatorname{supp}(\mathbf X_N)
\subseteq
\bigcup_{q=1}^{Q_c}\mathcal I_{r,q}\times\mathcal I_{t,q},
\label{eq:xN_cluster_support}
\end{equation}
which gives the block-sparse structure of the NLOS field.

For \(i\in\mathcal I_{r,q}\) and \(j\in\mathcal I_{t,q}\), define the antenna-level atom gates by \([\mathbf Z_{r,{\rm ant}}]_{n_r,i}=\nu_{r,q}^{A}(\boldsymbol\rho_{r,n_r})\) and \([\mathbf Z_{t,{\rm ant}}]_{n_t,j}=\nu_{t,q}^{A}(\boldsymbol\rho_{t,n_t})\). The antenna-level array manifolds are
\begin{align}
\mathbf A_{{\rm Rx},{\rm ant}}
&=\left[\mathbf a_{\rm Rx}^{T}(\boldsymbol\rho_{r,1}),\ldots,
\mathbf a_{\rm Rx}^{T}(\boldsymbol\rho_{r,N_r})\right]^T
\in\mathbb C^{N_r\times Q_r},
\notag\\[-0.2em]
\mathbf A_{{\rm Tx},{\rm ant}}
&=\left[\mathbf a_{\rm Tx}^{T}(\boldsymbol\rho_{t,1}),\ldots,
\mathbf a_{\rm Tx}^{T}(\boldsymbol\rho_{t,N_t})\right]^T
\in\mathbb C^{N_t\times Q_t},
\label{eq:antenna_level_manifolds}
\end{align}
where
\begin{align}
\mathbf a_{\rm Rx}(\boldsymbol\rho)
&=\frac{1}{\sqrt{N_r}}
\left[e^{\jj\boldsymbol\kappa_{r,1}^{T}\boldsymbol\rho},\ldots,
e^{\jj\boldsymbol\kappa_{r,Q_r}^{T}\boldsymbol\rho}\right],
\notag\\[-0.2em]
\mathbf a_{\rm Tx}(\boldsymbol\rho)
&=\frac{1}{\sqrt{N_t}}
\left[e^{-\jj\boldsymbol\kappa_{t,1}^{T}\boldsymbol\rho},\ldots,
e^{-\jj\boldsymbol\kappa_{t,Q_t}^{T}\boldsymbol\rho}\right].
\label{eq:antenna_level_responses}
\end{align}
The corresponding gated antenna-level dictionaries are \(\widetilde{\mathbf A}_{{\rm Rx},{\rm ant}}=\mathbf Z_{r,{\rm ant}}\odot\mathbf A_{{\rm Rx},{\rm ant}}\) and \(\widetilde{\mathbf A}_{{\rm Tx},{\rm ant}}=\mathbf Z_{t,{\rm ant}}\odot\mathbf A_{{\rm Tx},{\rm ant}}\). Thus, the NLOS field in \eqref{eq:discrete_Enlos_synthesis} can be arranged as
\begin{equation}
\mathbf E_{{\rm NLOS},{\rm ant}}
=\widetilde{\mathbf A}_{{\rm Rx},{\rm ant}}
\mathbf X_N
\widetilde{\mathbf A}_{{\rm Tx},{\rm ant}}^{T}
+\boldsymbol\varepsilon_N,
\label{eq:Enlos_antenna_matrix}
\end{equation}

Based on similar derivations from \eqref{eq:Elos_spatial_response} to \eqref{eq:discrete_Enlos_synthesis}, the LOS response is also sparse in the wavenumber domain. Specifically, the Green-function integral in \eqref{eq:Elos_spatial_response} is discretized over the LOS support sets \(\mathcal I_{r,L}\) and \(\mathcal I_{t,L}\), which are the index sets of the retained receive and transmit-side wavenumber atoms around the LOS stationary point, respectively. As a result, we have
\begin{align}
&E_{\mathcal A}^{\rm LOS}(\boldsymbol\rho_{r,n_r},\boldsymbol\rho_{t,n_t})
=\nu_L^A(\boldsymbol\rho_{r,n_r},\boldsymbol\rho_{t,n_t})\notag\\[-0.2em]
&\cdot\sum_{\ell\in\mathcal I_{r,L}}\sum_{j\in\mathcal I_{t,L}}x_{L,\ell,j}
e^{\jj\boldsymbol\kappa_{L,\ell}^{T}\boldsymbol\rho_{r,n_r}
-\jj\boldsymbol\kappa_{t,L,j}^{T}\boldsymbol\rho_{t,n_t}}
+ [\boldsymbol\varepsilon_L]_{n_r,n_t}.
\label{eq:discrete_Elos_synthesis}
\end{align}
\begingroup
\interlinepenalty=10000
Here, \(\mathcal I_{r,L}=\{\ell:\boldsymbol\kappa_{L,\ell}\in\mathcal N(\mathcal K_{r,L})\}\) and \(\mathcal I_{t,L}=\{j:\boldsymbol\kappa_{t,L,j}\in\mathcal N(\mathcal K_{t,L})\}\), with \(\mathcal K_{r,L}=\mathcal K_{t,L}=\allowbreak\bigl\{-k_c\frac{\boldsymbol\rho_r-\boldsymbol\rho_t}{\|\mathbf r_r(\boldsymbol\rho_r)-\mathbf r_t(\boldsymbol\rho_t)\|_2}:\allowbreak\boldsymbol\rho_r\in\mathcal A_r,\ \allowbreak\boldsymbol\rho_t\in\mathcal A_t\bigr\}\). The vectors \(\boldsymbol\kappa_{L,\ell}\) and \(\boldsymbol\kappa_{t,L,j}\) denote the receive- and transmit-side LOS wavenumber grid points, respectively.
\par\endgroup
Let \(\mathbf X_L=[x_{L,\ell,j}]\in\mathbb C^{Q_L\times Q_L}\), where \(Q_L\) is the number of retained LOS wavenumber atoms on each side. Define the LOS array manifolds \(\mathbf A_{L,{\rm ant}}\in\mathbb C^{N_r\times Q_L}\) and \(\mathbf A_{{\rm Tx,LOS},{\rm ant}}\in\mathbb C^{N_t\times Q_L}\), which have forms similar to \(\mathbf A_{{\rm Rx},{\rm ant}}\) and \(\mathbf A_{{\rm Tx},{\rm ant}}\), respectively, except that they are constructed from the LOS wavenumber grids. The receive-side LOS response is \(\mathbf a_L(\boldsymbol\rho)=N_r^{-1/2}[e^{\jj\boldsymbol\kappa_{L,\ell}^{T}\boldsymbol\rho}]_{\ell\in\mathcal I_{r,L}}\), with \([\mathbf A_{L,{\rm ant}}]_{n_r,:}=\mathbf a_L(\boldsymbol\rho_{r,n_r})\).

Similar to the NLOS model, where cluster visibility is converted to wavenumber-atom visibility, we likewise convert the LOS visibility to wavenumber-atom visibility and accordingly define \(\mathbf Z_{L,{\rm ant}}\in\{0,1\}^{N_r\times Q_L}\) and the corresponding gated antenna-level dictionary \(\widetilde{\mathbf A}_{L,{\rm ant}}=\mathbf Z_{L,{\rm ant}}\odot\mathbf A_{L,{\rm ant}}\). Thus,
\begin{equation}
\mathbf E_{{\rm LOS},{\rm ant}}
=\widetilde{\mathbf A}_{L,{\rm ant}}
\mathbf X_L\mathbf A_{{\rm Tx,LOS},{\rm ant}}^{T}
+\boldsymbol\varepsilon_L.
\label{eq:Elos_antenna_matrix}
\end{equation}
The nonzero elements of \(\mathbf X_L\) are concentrated at the paired receive--transmit wavenumber positions of the direct path and do not form clear two-dimensional blocks. Therefore, we absorb the transmit LOS dictionary into \(\mathbf X_L\) to get \(\mathbf W_{L,{\rm ant}}=\mathbf X_L\mathbf A_{{\rm Tx,LOS},{\rm ant}}^{T}\). The resulting \(\mathbf W_{L,{\rm ant}}\) is row sparse and can therefore be estimated together with the block-sparse NLOS coefficients.

We next obtain the grouped channel model. Let \(\boldsymbol\rho_{u,g_u}\) be the center of group \(g_u\), with \(g_u(n_u)\) denoting the group containing antenna \(n_u\), where \(u\in\{r,t\}\). When the group size is relatively small, the antennas within the same group can be assumed to share the same spatial response and visibility. Therefore, the spatial response and visibility of antenna \(n_u\) are approximated by those at \(\boldsymbol\rho_{u,g_u(n_u)}\).

Let \(\mathbf Z_r\in\{0,1\}^{G_r\times Q_r}\) and \(\mathbf Z_t\in\{0,1\}^{G_t\times Q_t}\) collect the receive- and transmit-group visibility of the retained atoms in the wavenumber domain. For atoms in the \(q\)-th NLOS block, \([\mathbf Z_r]_{g_r,i}=\nu_{r,q}^{A}(\boldsymbol\rho_{r,g_r})\) and \([\mathbf Z_t]_{g_t,j}=\nu_{t,q}^{A}(\boldsymbol\rho_{t,g_t})\), for \(i\in\mathcal I_{r,q}\) and \(j\in\mathcal I_{t,q}\). The LOS visibility is collected in \(\mathbf Z_L\in\{0,1\}^{G_r\times Q_L}\). Define the receive-group membership matrix as
\begin{equation}
[\mathbf M_r]_{n_r,g_r}
=
\begin{cases}
1,&g_r=g_r(n_r),\\
0,&\text{otherwise},
\end{cases}
\label{eq:receive_membership_matrix}
\end{equation}
The group-center dictionaries are \([\mathbf A_{{\rm Rx},G}]_{g_r,:}=\mathbf a_{\rm Rx}(\boldsymbol\rho_{r,g_r})\), \([\mathbf A_{{\rm Tx},G}]_{g_t,:}=\eta_{g_t}\mathbf a_{\rm Tx}(\boldsymbol\rho_{t,g_t})\), and \([\mathbf A_{L,G}]_{g_r,:}=\mathbf a_L(\boldsymbol\rho_{r,g_r})\), where \(\eta_{g_t}=\sum_{n_t\in\mathcal U^t_{g_t}}\beta_{n_t,g_t}\). Under the group approximation, the NLOS and LOS fields at the receive antennas are given by
\begin{equation}
\begin{aligned}
&\mathbf E_{\rm NLOS}=\mathbf E_{{\rm NLOS},{\rm ant}}\mathbf B_t
=(\mathbf M_r\mathbf Z_r\odot\mathbf M_r\mathbf A_{{\rm Rx},G})\mathbf X_N\\[-0.2em]
&\quad\cdot(\mathbf Z_t\odot\mathbf A_{{\rm Tx},G})^{T}
+\boldsymbol\Delta_E^{\rm NLOS},\\[-0.2em]
&\mathbf E_{\rm LOS}=\mathbf E_{{\rm LOS},{\rm ant}}\mathbf B_t
=(\mathbf M_r\mathbf Z_L\odot\mathbf M_r\mathbf A_{L,G})\mathbf W_L
+\boldsymbol\Delta_E^{\rm LOS}.
\end{aligned}
\label{eq:group_center_fields}
\end{equation}
where \(\mathbf W_L=\mathbf W_{L,{\rm ant}}\mathbf B_t\) and \([\mathbf B_t]_{n_t,g_t}=\beta_{n_t,g_t}\). The residual matrices are \(\boldsymbol\Delta_E^{\rm NLOS}=\widetilde{\mathbf A}_{{\rm Rx},{\rm ant}}\mathbf X_N\widetilde{\mathbf A}_{{\rm Tx},{\rm ant}}^T\mathbf B_t\allowbreak-(\mathbf M_r\mathbf Z_r\odot\mathbf M_r\mathbf A_{{\rm Rx},G})\mathbf X_N(\mathbf Z_t\odot\mathbf A_{{\rm Tx},G})^T\allowbreak+\boldsymbol\varepsilon_N\mathbf B_t\) and \(\boldsymbol\Delta_E^{\rm LOS}=\bigl[\widetilde{\mathbf A}_{L,{\rm ant}}-(\mathbf M_r\mathbf Z_L\odot\mathbf M_r\mathbf A_{L,G})\bigr]\mathbf W_L\allowbreak+\boldsymbol\varepsilon_L\mathbf B_t\). The row sparsity of \(\mathbf W_{L,{\rm ant}}\) makes \(\mathbf W_L\) a row-sparse matrix as well. The sum of the two terms excluding the residuals in \eqref{eq:group_center_fields} is denoted by \(\mathbf E_{\rm mod}\).

The actual equivalent electric field with grouped pilots is
\begin{equation}
\mathbf E_{\rm act}
=\mathbf E_{\rm LOS}+\mathbf E_{\rm NLOS}
=\mathbf E_{\rm mod}+\boldsymbol\Delta_E,
\label{eq:actual_group_field}
\end{equation}
where \(\boldsymbol\Delta_E=\boldsymbol\Delta_E^{\rm LOS}+\boldsymbol\Delta_E^{\rm NLOS}\).

Based on the analysis in Sec.~III-A, the space-time response field at \(\mathbf r_{n_r}\) due to excitation from the \(g_t\)-th transmit group is
\begin{equation}
e(\mathbf r_{n_r},g_t,t)
= [\mathbf E_{\rm act}]_{n_r,g_t}\widetilde i_{g_t}(t)e^{\jj\omega_ct}.
\label{eq:group_field_time_response}
\end{equation}
The receive-port conversion is represented by a calibrated diagonal matrix \(\mathbf C_{\rm port}\), which converts the sampled electric field to the received current signal. The resulting effective channel is
\begin{equation}
\begin{aligned}
\mathbf H_{\rm eff}&=\mathbf C_{\rm port}\mathbf E_{\rm act}
=\mathbf H_{\rm mod}+\mathbf C_{\rm port}\boldsymbol\Delta_E,\\[-0.2em]
\mathbf H_{\rm mod}&=\mathbf C_{\rm port}\mathbf E_{\rm mod}.
\end{aligned}
\label{eq:field_to_channel}
\end{equation}
The matrix representation of the received signal is given by
\begin{equation}
\mathbf Y=\mathbf H_{\rm eff}\mathbf S+\mathbf N_{\rm rx}
=\mathbf H_{\rm mod}\mathbf S+\mathbf N_{\rm eff},
\label{eq:Y_grouped_matrix}
\end{equation}
where \(\mathbf N_{\rm eff}=\mathbf C_{\rm port}\boldsymbol\Delta_E\mathbf S+\mathbf N_{\rm rx}\) includes the model errors and measurement noise.

At the receiver, the signals from the same group of receiving antennas are averaged to obtain the grouped observation signal. Define \(\mathbf D_r=\diag(|\mathcal U^r_1|,\ldots,|\mathcal U^r_{G_r}|)\) and \(\mathbf R_r=\mathbf D_r^{-1/2}\mathbf M_r^T\). The observation after grouping is
\begin{equation}
\mathbf Y_G
=\mathbf R_r\mathbf Y
=\mathbf H_{{\rm mod},G}\mathbf S+\mathbf N_{{\rm eff},G},
\label{eq:Y_receive_grouped}
\end{equation}
where \(\mathbf H_{{\rm mod},G}=\mathbf R_r\mathbf H_{\rm mod}\) and \(\mathbf N_{{\rm eff},G}=\mathbf R_r\mathbf N_{\rm eff}=\mathbf R_r\mathbf C_{\rm port}\boldsymbol\Delta_E\mathbf S+\mathbf R_r\mathbf N_{\rm rx}\).
Based on \eqref{eq:group_center_fields} and \eqref{eq:field_to_channel}, we can get
\begin{equation}
\mathbf H_{{\rm mod},G}
=\underbrace{\widetilde{\mathbf A}_{L,G}(\mathbf Z_L)\mathbf W_L}
_{\mathbf H_{{\rm mod},G}^{\rm LOS}}
+\underbrace{\widetilde{\mathbf A}_{{\rm Rx},G}(\mathbf Z_r)
\mathbf X_N
\widetilde{\mathbf A}_{\rm Tx}^{T}(\mathbf Z_t)}
_{\mathbf H_{{\rm mod},G}^{\rm NLOS}},
\label{eq:Hmod_grouped_explicit}
\end{equation}
where \(\widetilde{\mathbf A}_{{\rm Rx},G}(\mathbf Z_r)=\mathbf R_r\mathbf C_{\rm port}\mathbf M_r(\mathbf Z_r\odot\mathbf A_{{\rm Rx},G})\), \(\widetilde{\mathbf A}_{\rm Tx}(\mathbf Z_t)=\mathbf Z_t\odot\mathbf A_{{\rm Tx},G}\), and \(\widetilde{\mathbf A}_{L,G}(\mathbf Z_L)=\mathbf R_r\mathbf C_{\rm port}\mathbf M_r(\mathbf Z_L\odot\mathbf A_{L,G})\) are the gated group-center dictionaries.

\section{Visibility-Aware Block-Sparse Bayesian Estimation}
\label{sec:sampled_grid_model}

This section estimates the grouped channel $\mathbf H_{{\rm mod},G}$ in \eqref{eq:Hmod_grouped_explicit} from the observation in \eqref{eq:Y_receive_grouped}.
Due to the block sparsity and spatial correlation of the channel, we can obtain the channel estimate by recovering the sparse matrices \((\mathbf W_L,\mathbf X_N)\) and visibility matrices \((\mathbf Z_L,\mathbf Z_r,\mathbf Z_t)\).

\subsection{Bidirectional Visibility-Aware Observation Model}

The observation model includes scattering-cluster visibility at both the transmitter and the receiver. For receive wavenumber atom \(i\), transmit wavenumber atom \(j\), and LOS atom \(\ell\), the corresponding visibility vectors are
\(\mathbf z_i^r=[z_{1,i}^r,\ldots,z_{G_r,i}^r]^T=[\mathbf Z_r]_{:,i}\),
\(\mathbf z_j^t=[z_{1,j}^t,\ldots,z_{G_t,j}^t]^T=[\mathbf Z_t]_{:,j}\), and
\(\mathbf z_\ell^L=[z_{1,\ell}^L,\ldots,z_{G_r,\ell}^L]^T=[\mathbf Z_L]_{:,\ell}\).
The coefficient matrices \(\mathbf W_L\) and \(\mathbf X_N\) are shared by all groups, whereas the visibility matrices determine the gated dictionaries \(\widetilde{\mathbf A}_{{\rm Rx},G}(\mathbf Z_r)\), \(\widetilde{\mathbf A}_{\rm Tx}(\mathbf Z_t)\), and \(\widetilde{\mathbf A}_{L,G}(\mathbf Z_L)\).

Let
\(\boldsymbol\theta=[\vecop(\mathbf W_L)^T,\vecop(\mathbf X_N)^T]^T\),
\(\mathbf Z=(\mathbf Z_L,\mathbf Z_r,\mathbf Z_t)\),
\(\mathbf n_{{\rm eff},G}=\vecop(\mathbf N_{{\rm eff},G})\),
and
\(\mathbf y_G=\vecop(\mathbf Y_G)\).
We model
\(\mathbf n_{{\rm eff},G}\sim\mathcal{CN}(\mathbf0,\lambda\mathbf I_{G_rP})\), where \(\lambda\) is the parameter to be learned. Then
\begin{equation}
\mathbf y_G
=
\boldsymbol\Phi_G(\mathbf Z)\boldsymbol\theta
+
\mathbf n_{{\rm eff},G},
\label{eq:vec_obs_bidir}
\end{equation}
where
\begin{equation}
\boldsymbol\Phi_G(\mathbf Z)
=
\Big[
\mathbf S^T\otimes\widetilde{\mathbf A}_{L,G}(\mathbf Z_L),
\;
\big(\mathbf S^T\widetilde{\mathbf A}_{\rm Tx}(\mathbf Z_t)\big)
\otimes
\widetilde{\mathbf A}_{{\rm Rx},G}(\mathbf Z_r)
\Big].
\label{eq:explicit_sensing_matrix}
\end{equation}
Let \(\mathbf y_{G,g_r}=[\mathbf Y_G]_{g_r,:}^{T}\in\mathbb C^P\), and \(\boldsymbol\Phi_{G,g_r}\) collect the corresponding \(P\) rows of \(\boldsymbol\Phi_G\). Then, the likelihood is
\begin{equation}
p(\mathbf y_G\mid\boldsymbol\theta,\mathbf Z,\lambda)
=
\prod_{g_r=1}^{G_r}
\mathcal{CN}\!\left(
\mathbf y_{G,g_r};
\boldsymbol\Phi_{G,g_r}(\mathbf Z)\boldsymbol\theta,
\lambda\mathbf I_P
\right).
\label{eq:groupwise_likelihood}
\end{equation}

\subsection{Bidirectional Markov Visibility Prior}

We arrange the 2D antenna groups along a side-adjacent serpentine path, so that in each \(\mathbf z_i^r\), \(\mathbf z_i^t\), or \(\mathbf z_i^L\), adjacent elements correspond to spatially neighboring groups. The visibility of spatially adjacent groups to the same cluster exhibits strong correlation. Therefore, the visibility sequence of each wavenumber atom is modeled by a first-order Markov chain. For \(u\in\{r,t,L\}\), \(a,b\in\{0,1\}\), and any atom \(i\), \(T^u_{a,b}=p(z_{g,i}^u=b\mid z_{g-1,i}^u=a)\) is the conditional probability of adjacent visibility variables. The conditional probability matrix is
\begin{equation}
\mathbf T^u=
\begin{bmatrix}
T_{0,0}^u&T_{0,1}^u\\
T_{1,0}^u&T_{1,1}^u
\end{bmatrix}.
\label{eq:transition_matrix_main}
\end{equation}
Using the entries of \(\mathbf T^u\), for any \(u\in\{r,t,L\}\), the prior for the visibility vector \(\mathbf z_i^u\) is given by
\begin{align}
p(\mathbf z_i^u)
&=p(z_{1,i}^u)\prod_{g=2}^{G_u}
p(z_{g,i}^u\mid z_{g-1,i}^u),
\label{eq:markov_prior_main}
\end{align}
where \(G_u=G_r\) for \(u\in\{r,L\}\) and \(G_u=G_t\) for \(u=t\).
Based on the steady-state probability analysis of a first-order Markov chain, we can obtain
\begin{equation}
p(z_{1,i}^u=1)=\frac{T_{0,1}^u}{T_{0,1}^u+T_{1,0}^u},\quad
p(z_{1,i}^u=0)=\frac{T_{1,0}^u}{T_{0,1}^u+T_{1,0}^u}.
\label{eq:markov_stationary_initial}
\end{equation}

\subsection{Block-Sparse Bayesian Coefficient and Visibility Estimation}
\label{sec:sbl_estimator}

The sparse vector \(\boldsymbol\theta\) and visibility variables \(\mathbf Z\) are updated alternately. First, we estimate \(\boldsymbol\theta\) with fixed \(\mathbf Z\). Then, \(\mathbf Z\) is updated conditioned on the estimate of \(\boldsymbol\theta\), followed by a restricted coefficient refit.

\subsubsection{Block-Sparse Coefficient Update With Fixed Visibility}

For fixed \(\mathbf Z\), divide
\(\boldsymbol\theta\) into LOS-row and NLOS-block groups \(\{\mathcal G_b\}_{b=1}^{B}\). For each group, assume
\(\boldsymbol\theta_{\mathcal G_b}\sim\mathcal{CN}(\mathbf0,\gamma_b\mathbf I)\), and groups are independent. The inverse variances have independent Gamma priors, \(\gamma_b^{-1}\sim\operatorname{Gamma}(a_\gamma,b_\gamma)\), with fixed small positive shape and rate parameters \(a_\gamma\) and \(b_\gamma\), respectively, taken in the limit \(a_\gamma,b_\gamma\to0^+\). Thus, the effective prior covariance \(\boldsymbol\Gamma(\boldsymbol\gamma)\) is diagonal, with \([\boldsymbol\Gamma(\boldsymbol\gamma)]_{kk}=\gamma_b\) for \(k\in\mathcal G_b\).
The variational posterior covariance and mean of \(\boldsymbol\theta\) are
\begin{align}
\boldsymbol\Sigma_\theta
&=\left(
\boldsymbol\Gamma^{-1}(\boldsymbol\gamma)
+\lambda^{-1}\boldsymbol\Phi_G^H(\mathbf Z)
\boldsymbol\Phi_G(\mathbf Z)
\right)^{-1},
\label{eq:Sigma_theta}\\[-0.2em]
\boldsymbol\mu_\theta
&=\lambda^{-1}\boldsymbol\Sigma_\theta
\boldsymbol\Phi_G^H(\mathbf Z)\mathbf y_G .
\label{eq:mu_theta}
\end{align}
Let \(\boldsymbol\mu_{\mathcal G_b}\) and \(\boldsymbol\Sigma_{\mathcal G_b,\mathcal G_b}\) denote the posterior mean entries and covariance submatrix associated with group \(\mathcal G_b\) of \(\boldsymbol\theta\). Using the reciprocals of the posterior mean precisions for \(\boldsymbol\gamma\) and maximizing the expected log-likelihood over \(\lambda\), we obtain
\begin{align}
\gamma_b^{\rm new}
&=\frac{b_\gamma+\|\boldsymbol\mu_{\mathcal G_b}\|_2^2
+\tr(\boldsymbol\Sigma_{\mathcal G_b,\mathcal G_b})}
{a_\gamma+|\mathcal G_b|},
\qquad b=1,\ldots,B,
\label{eq:block_gamma_update}\\[-0.2em]
\lambda^{\rm new}
&=\frac{
\|\mathbf y_G-\boldsymbol\Phi_G(\mathbf Z)\boldsymbol\mu_\theta\|_2^2
+\tr\!\left(
\boldsymbol\Phi_G(\mathbf Z)
\boldsymbol\Sigma_\theta
\boldsymbol\Phi_G^H(\mathbf Z)
\right)}
{G_rP}.
\label{eq:lambda_update}
\end{align}
The iteration stops when the maximum of \(\|\boldsymbol\gamma^{\rm new}-\boldsymbol\gamma\|_2/\|\boldsymbol\gamma\|_2\) and \(|\lambda^{\rm new}-\lambda|/|\lambda|\) is below \(\epsilon\). After convergence, the posterior mean is reshaped into
\(\widehat{\mathbf W}_L\) and \(\widehat{\mathbf X}_N\).

Based on \(\widehat{\mathbf W}_L\) and \(\widehat{\mathbf X}_N\), the candidate LOS rows and NLOS blocks are selected and the remaining LOS and NLOS entries are set to zero to form \(\widehat{\mathbf W}_L^{\rm cur}\) and \(\widehat{\mathbf X}_N^{\rm cur}\). For the visibility update, set \(\widehat{\mathbf W}_L\leftarrow\widehat{\mathbf W}_L^{\rm cur}\) and \(\widehat{\mathbf X}_N\leftarrow\widehat{\mathbf X}_N^{\rm cur}\). Let \(\mathcal A_L^{\rm cand}\) contain the selected LOS row indices and \(\widehat{\mathcal S}_N^{\rm cand}\) contain the coefficient index pairs in the selected NLOS blocks, including the one-ring neighbors of strong blocks. The corresponding receive and transmit atom sets are \(\mathcal I_r^{\rm cand}=\{i:\exists j,\,(i,j)\in\widehat{\mathcal S}_N^{\rm cand}\}\) and \(\mathcal I_t^{\rm cand}=\{j:\exists i,\,(i,j)\in\widehat{\mathcal S}_N^{\rm cand}\}\), respectively.

\subsubsection{Low-Complexity Visibility Update}

Based on the likelihood in \eqref{eq:groupwise_likelihood} and the Markov prior in \eqref{eq:markov_prior_main}, we adopt the maximum a posteriori (MAP) criterion to jointly estimate the NLOS receive visibility \(\mathbf Z_r\), NLOS transmit visibility \(\mathbf Z_t\), and LOS visibility \(\mathbf Z_L\). Marginalizing \(\boldsymbol\theta\) under the prior specified in Section~IV-C1 gives
\begin{equation}
\widehat{\mathbf Z}
=\argmax_{\mathbf Z}
\left[-\log\det\mathbf C_{\mathbf Z}
-\mathbf y_G^H\mathbf C_{\mathbf Z}^{-1}\mathbf y_G
+\log p(\mathbf Z)\right],
\label{eq:visibility_marginal_objective}
\end{equation}
where \(\mathbf C_{\mathbf Z}=\lambda\mathbf I_{G_rP}+\boldsymbol\Phi_G(\mathbf Z)\boldsymbol\Gamma(\boldsymbol\gamma)\boldsymbol\Phi_G^H(\mathbf Z)\). However, solving \eqref{eq:visibility_marginal_objective} requires repeated inversion of the high-dimensional matrix \(\mathbf C_{\mathbf Z}\). As a result, we model \(\boldsymbol\theta\) as a deterministic quantity and set \(\boldsymbol\theta=[\vecop(\widehat{\mathbf W}_L)^T,\vecop(\widehat{\mathbf X}_N)^T]^T\). In this case, the MAP problem can be reformulated as
\begin{equation}
\widehat{\mathbf Z}
=\argmax_{\mathbf Z}
\left[-\|\mathbf E(\mathbf Z)\|_F^2
+\lambda\log p(\mathbf Z)\right],
\label{eq:visibility_conditional_objective}
\end{equation}
where \(\mathbf E(\mathbf Z)=\mathbf Y_G-\widehat{\mathbf H}_G(\mathbf Z)\mathbf S\) and \(\widehat{\mathbf H}_G(\mathbf Z)=\widetilde{\mathbf A}_{L,G}(\mathbf Z_L)\widehat{\mathbf W}_L^{\rm cur}+\widetilde{\mathbf A}_{{\rm Rx},G}(\mathbf Z_r)\widehat{\mathbf X}_N^{\rm cur}\widetilde{\mathbf A}_{\rm Tx}^{T}(\mathbf Z_t)\). Note that only the values in \(\mathbf Z\) that correspond to the nonzero entries of the sparse matrix \(\widehat{\mathbf W}_L^{\rm cur}\) or \(\widehat{\mathbf X}_N^{\rm cur}\) need to be re-optimized. Thus, optimizing \(\mathbf Z\) is equivalent to further sparsifying \(\widehat{\mathbf W}_L^{\rm cur}\) and \(\widehat{\mathbf X}_N^{\rm cur}\) by exploiting spatial correlation.
We solve \eqref{eq:visibility_conditional_objective} by updating \(\mathbf Z_t\), \(\mathbf Z_r\), and \(\mathbf Z_L\) alternately.

For the \(i\)-th receive atom, decompose \(\mathbf Z_r\) as \(\mathbf Z_r=\mathbf Z_{r,-i}+\mathbf z_i^r\mathbf e_i^T\), where \(\mathbf Z_{r,-i}\) is obtained by zeroing the \(i\)-th column of \(\mathbf Z_r\) and \(\mathbf e_i\) is the \(i\)-th canonical vector. Fixing \(\mathbf Z_t\), \(\mathbf Z_L\), and \(\mathbf Z_{r,-i}\) leaves only \(\mathbf z_i^r\) to be updated. Let \(\mathbf a_i^r=\mathbf R_r\mathbf C_{\rm port}\mathbf M_r[\mathbf A_{{\rm Rx},G}]_{:,i}\), \(a_{g_r,i}^r=[\mathbf a_i^r]_{g_r}\), and \(\mathbf b_i^r=[\widehat{\mathbf X}_N^{\rm cur}\widetilde{\mathbf A}_{\rm Tx}^{T}(\mathbf Z_t)\mathbf S]_{i,:}\). Removing the terms of all other atoms from \(\mathbf E(\mathbf Z)\) gives \(\mathbf R_{-i}^r=\mathbf E(\mathbf Z)+(\mathbf z_i^r\odot\mathbf a_i^r)\mathbf b_i^r\).

The squared residual is separable across receive groups, and the fitting cost of each group is
\begin{equation}
d_{i,g_r}^r(s)=\left\|[\mathbf R_{-i}^r]_{g_r,:}-s\,a_{g_r,i}^r\mathbf b_i^r\right\|_2^2,
\quad s\in\{0,1\}.
\label{eq:receive_emission_main}
\end{equation}
Consequently, the receive-side update is
\begingroup
\setlength{\abovedisplayskip}{4pt plus 1pt minus 1pt}
\setlength{\belowdisplayskip}{4pt plus 1pt minus 1pt}
\begin{equation}
\widehat{\mathbf z}_i^r
=\argmin_{\mathbf z_i^r\in\{0,1\}^{G_r}}
\left(
\sum_{g_r=1}^{G_r}
d_{i,g_r}^r(z_{g_r,i}^r)
-\lambda\log p(\mathbf z_i^r)
\right).
\label{eq:z_update_map_explicit}
\end{equation}
\endgroup
Because \(\mathbf z_i^r\) follows a first-order Markov process, \eqref{eq:z_update_map_explicit} is solved by a two-state Viterbi recursion. Let \(V_{i,g_r}^r(s)\) be the minimum accumulated cost up to group \(g_r\) with final state \(s\). Initialize \(V_{i,1}^r(s)=d_{i,1}^r(s)-\lambda\log p(z_{1,i}^r=s)\). For \(g_r=2,\ldots,G_r\), compute
\begin{equation}
V_{i,g_r}^r(s)=d_{i,g_r}^r(s)+\min_{s'\in\{0,1\}}
\left[V_{i,g_r-1}^r(s')-\lambda\log T_{s',s}^r\right].
\label{eq:viterbi_recursion_main}
\end{equation}
The minimizing predecessor of each state is stored as
\begin{equation}
\psi_{i,g_r}^r(s)=\argmin_{s'\in\{0,1\}}
\left[V_{i,g_r-1}^r(s')-\lambda\log T_{s',s}^r\right].
\label{eq:viterbi_predecessor_main}
\end{equation}
For each state, the lower of the two accumulated costs is retained, together with its predecessor \(\psi_{i,g_r}^r(s)\). After the forward recursion, the last state and the remaining states are recovered by
\begin{equation}
\begin{aligned}
\widehat z_{G_r,i}^{r}
&=\argmin_{s\in\{0,1\}}V_{i,G_r}^{r}(s),\\
\widehat z_{g_r-1,i}^{r}
&=\psi_{i,g_r}^{r}\!\left(\widehat z_{g_r,i}^{r}\right),
\quad g_r=G_r,\ldots,2,\\
\widehat{\mathbf z}_i^r
&=[\widehat z_{1,i}^{r},\ldots,\widehat z_{G_r,i}^{r}]^T.
\end{aligned}
\label{eq:app_receive_traceback_update}
\end{equation}

For fixed \(\widehat{\mathbf W}_L^{\rm cur}\), \(\widehat{\mathbf X}_N^{\rm cur}\), \(\mathbf Z_L\), \(\mathbf Z_t\), and \(\mathbf Z_{r,-i}\), the cost in \eqref{eq:z_update_map_explicit} contains only the single-group costs \(d_{i,g_r}^r\) and the transition cost between two adjacent states. Any sequence ending in state \(s\) must therefore reach it from either \(s'=0\) or \(s'=1\). Minimizing the prefix first gives exactly the recursion in \eqref{eq:viterbi_recursion_main}. In this way, dynamic programming accounts for all \(2^{G_r}\) binary sequences without listing them one by one, and the traceback returns the global minimizer of the conditional subproblem for \(\mathbf z_i^r\).

For the NLOS transmit visibility, \(\mathbf S\) mixes the contributions from different transmit groups. When \(\mathbf S\mathbf S^H\) is not diagonal, the fitting error generally contains cross terms between their visibility states and cannot be separated as in \eqref{eq:z_update_map_explicit}; hence, the receive-side two-state Viterbi recursion cannot be applied directly. For the \(j\)-th retained transmit atom, we instead visit \(g_t=1,\ldots,G_t\) in order. Fixing all other states at their latest values, compute the objective value in \eqref{eq:visibility_conditional_objective} for state 1 minus that for state 0 as
\begin{equation*}
\begin{aligned}
\Delta_{g_t,j}^{t}
={}&\left.\|\mathbf E(\mathbf Z)\|_F^2\right|_{z_{g_t,j}^{t}=0}
-\left.\|\mathbf E(\mathbf Z)\|_F^2\right|_{z_{g_t,j}^{t}=1}\\
&+\lambda\left[
\left.\log p(\mathbf Z)\right|_{z_{g_t,j}^{t}=1}
-\left.\log p(\mathbf Z)\right|_{z_{g_t,j}^{t}=0}
\right].
\end{aligned}
\end{equation*}
The state is then updated by
\begin{equation}
\widehat z_{g_t,j}^{t}
=\begin{cases}
1, & \Delta_{g_t,j}^{t}>0,\\
0, & \Delta_{g_t,j}^{t}<0,\\
z_{g_t,j}^{t}, & \Delta_{g_t,j}^{t}=0.
\end{cases}
\label{eq:transmit_visibility_update}
\end{equation}
Although the transmit-side fitting error cannot be separated as in \eqref{eq:z_update_map_explicit}, we can still reduce the computational cost by exploiting the sparsity of \(\widehat{\mathbf X}_N^{\rm cur}\). Specifically, \eqref{eq:transmit_visibility_update} is applied only to retained transmit atoms with nonzero coefficient columns.

For the LOS visibility, each state also affects only one receive row, so the fitting error separates across receive groups as in \eqref{eq:z_update_map_explicit}. For LOS atom \(\ell\), let \(\mathbf a_\ell^L=\mathbf R_r\mathbf C_{\rm port}\mathbf M_r[\mathbf A_{L,G}]_{:,\ell}\), \(a_{g_r,\ell}^L=[\mathbf a_\ell^L]_{g_r}\), and \(\mathbf b_\ell^L=[\widehat{\mathbf W}_L^{\rm cur}]_{\ell,:}\mathbf S\). Adding back its current contribution gives \(\mathbf R_{-\ell}^L=\mathbf E(\mathbf Z)+(\mathbf z_\ell^L\odot\mathbf a_\ell^L)\mathbf b_\ell^L\). With the group cost \(d_{\ell,g_r}^L(s)=\|[\mathbf R_{-\ell}^L]_{g_r,:}-s\,a_{g_r,\ell}^L\mathbf b_\ell^L\|_2^2\), \eqref{eq:visibility_conditional_objective} yields
\begin{equation}
\widehat{\mathbf z}_\ell^L
=\argmin_{\mathbf z_\ell^L\in\{0,1\}^{G_r}}
\left[\sum_{g_r=1}^{G_r}d_{\ell,g_r}^L(z_{g_r,\ell}^L)
-\lambda\log p(\mathbf z_\ell^L)\right].
\label{eq:los_visibility_main}
\end{equation}
This update also uses the two-state Viterbi algorithm. Let \(V_{\ell,g_r}^L(s)\) and \(\psi_{\ell,g_r}^L(s)\) denote the minimum accumulated cost and the minimizing predecessor at group \(g_r\), respectively. Initialize \(V_{\ell,1}^L(s)=d_{\ell,1}^L(s)-\lambda\log p(z_{1,\ell}^L=s)\), and apply \eqref{eq:viterbi_recursion_main} and \eqref{eq:viterbi_predecessor_main} with \((r,i)\) replaced by \((L,\ell)\) and \(\mathbf T^r\) by \(\mathbf T^L\). The LOS visibility sequence is then recovered by
\begin{equation}
\begin{aligned}
\widehat z_{G_r,\ell}^{L}
&=\argmin_{s\in\{0,1\}}V_{\ell,G_r}^{L}(s),\\
\widehat z_{g_r-1,\ell}^{L}
&=\psi_{\ell,g_r}^{L}\!\left(\widehat z_{g_r,\ell}^{L}\right),
\quad g_r=G_r,\ldots,2,\\
\widehat{\mathbf z}_\ell^L
&=[\widehat z_{1,\ell}^{L},\ldots,\widehat z_{G_r,\ell}^{L}]^T.
\end{aligned}
\label{eq:los_visibility_traceback_update}
\end{equation}

Algorithm~\ref{alg:proposed_sbl} summarizes the alternating updates. With \(\mathbf Z\) fixed, repeat steps~\ref{algline:sbl}--\ref{algline:hyper} to the SBL stopping tolerance \(\epsilon\), then perform step~\ref{algline:screen}. The initial SBL run uses all coefficients; later refits use only the retained candidates. During each visibility round, keep the coefficients, \(\lambda\), and candidate sets fixed. Repeat step~\ref{algline:receive} until \(\mathbf Z_r\) is unchanged, and repeat steps~\ref{algline:receive}--\ref{algline:los} until \(\mathbf Z\) is unchanged. A round that changes \(\mathbf Z\) is followed by a coefficient refit through steps~\ref{algline:sbl}--\ref{algline:screen}, including the last round. Stop if a round leaves \(\mathbf Z\) unchanged or after \(T_z\) visibility rounds. Only retained atoms whose coefficient norms exceed \(\tau_{\rm vis}\) are eligible: use row norms for NLOS receive and LOS atoms, and column norms for NLOS transmit atoms. Update eligible sequences and states sequentially using their latest values. Form \(\mathbf E(\mathbf Z)\) at the start of each round and refresh it after every sequence or state update. Retain the current state or sequence if it is also optimal. Before reconstruction, zero LOS rows with \(\ell_2\) norm below \(\tau_{L,f}\) and NLOS blocks with Frobenius norm below \(\tau_f\).

\begin{algorithm}[t]
\caption{Visibility-aware block-SBL channel estimation}
\label{alg:proposed_sbl}
\small
\setlength{\leftmargini}{1.5em}
\setlength{\leftmarginii}{1.3em}
\renewcommand{\labelenumi}{\arabic{enumi})}
\renewcommand{\labelenumii}{\alph{enumii})}
\renewcommand{\theenumii}{(\alph{enumii})}
\begin{enumerate}
\setlength{\itemsep}{0pt}
\setlength{\parsep}{0pt}
\setlength{\topsep}{1pt}
\item \textbf{Input:} \(\mathbf Y_G\) and \(\mathbf S\).
\item \textbf{Initialization:} \(\mathbf Z_u\leftarrow\mathbf1\), \(u\in\{L,r,t\}\); \(\boldsymbol\gamma>\mathbf0\) and \(\lambda>0\).
\item Repeat the following until convergence:
\begin{enumerate}
\setlength{\itemsep}{0pt}
\setlength{\parsep}{0pt}
\setlength{\topsep}{0pt}
\item\label{algline:sbl} Update \(\boldsymbol\Sigma_\theta\) and \(\boldsymbol\mu_\theta\) with \eqref{eq:Sigma_theta} and \eqref{eq:mu_theta}, respectively.
\item\label{algline:hyper} Update \(\gamma_b\) and \(\lambda\) with \eqref{eq:block_gamma_update} and \eqref{eq:lambda_update}, respectively, \(\forall b\).
\item\label{algline:screen} Update \(\widehat{\mathbf W}_L\) and \(\widehat{\mathbf X}_N\) as in Section~IV-C1.
\item\label{algline:receive} Update \(\mathbf z_i^r\) with \eqref{eq:app_receive_traceback_update}, \(\forall i\).
\item\label{algline:transmit} Update \(z_{g_t,j}^t\) with \eqref{eq:transmit_visibility_update}, \(\forall g_t,j\).
\item\label{algline:los} Update \(\mathbf z_\ell^L\) with \eqref{eq:los_visibility_traceback_update}, \(\forall\ell\).
\end{enumerate}
\item\label{algline:reconstruct} \textbf{Output:} \(\widehat{\mathbf H}_G\) from \eqref{eq:Hmod_grouped_explicit}, \(\widehat{\mathbf W}_L\), \(\widehat{\mathbf X}_N\), and \(\widehat{\mathbf Z}=\mathbf Z\).
\end{enumerate}
\end{algorithm}

\subsubsection{Hybrid CRWW Performance Bound}
\label{subsec:hybrid_bound}

To obtain a performance reference for channel estimation, we start from the grouped observations in \eqref{eq:Y_receive_grouped}. For this reference, the array geometry is known. The complete observation model can be written as
\begin{align}
[\mathbf Y_G]_{g_r,p}
={}&\sum_{n_r=1}^{N_r}\sum_{n_t=1}^{N_t}
[\mathbf R_r\mathbf C_{\rm port}]_{g_r,n_r}
\Bigl[E_{\mathcal A}^{\rm LOS}(\boldsymbol\rho_{r,n_r},\boldsymbol\rho_{t,n_t})\notag\\[-0.2em]
&+E_{\mathcal A}^{\rm NLOS}(\boldsymbol\rho_{r,n_r},\boldsymbol\rho_{t,n_t})\Bigr]
[\mathbf B_t\mathbf S]_{n_t,p}
+[\mathbf R_r\mathbf N_{\rm rx}]_{g_r,p}\notag\\[-0.2em]
={}&[\mathbf H_G(\boldsymbol\vartheta,\mathbf z)\mathbf S
+\mathbf R_r\mathbf N_{\rm rx}]_{g_r,p}.
\label{eq:hybrid_complete_observation}
\end{align}
where the aperture coordinates in each summand satisfy \(\mathbf r_r(\boldsymbol\rho_{r,n_r})=\mathbf r_{n_r}\) and \(\mathbf r_t(\boldsymbol\rho_{t,n_t})=\mathbf r_{n_t}\). The fields \(E_{\mathcal A}^{\rm LOS}\) and \(E_{\mathcal A}^{\rm NLOS}\) are evaluated using only the continuous Green terms in \eqref{eq:Elos_aperture_continuous} and \eqref{eq:Enlos_aperture_continuous}, respectively, excluding the field-modeling residuals \(e_{\rm L}\) and \(e_{\rm N}\).
The receiver-noise variance \(\lambda_n\) is specified by the received signal power and SNR, rather than estimated by the algorithm, and is held fixed when evaluating the bound.
The continuous parameters \(\boldsymbol\vartheta=\{\alpha_L,(\alpha_q,\mathbf c_q)_{q=1}^{Q_c}\}\) are deterministic but unknown, with no prior assigned. The discrete parameters \(\mathbf z=\{\nu_L^A(\boldsymbol\rho_{r,g_r},\boldsymbol\rho_{t,g_t}),\nu_{r,q}^A(\boldsymbol\rho_{r,g_r}),\nu_{t,q}^A(\boldsymbol\rho_{t,g_t})\}\) follow the same first-order Markov form as \eqref{eq:markov_prior_main}--\eqref{eq:markov_stationary_initial}. The LOS receive sequence for each transmit group follows this prior form with \(\mathbf T^L\). These sequences are mutually independent and independent of \(\boldsymbol\vartheta\), giving the joint prior \(p_{\rm M}(\mathbf z)\). The binary visibility states do not admit ordinary parameter derivatives. We therefore use the hybrid Cram\'er--Rao--Weiss--Weinstein bound (HCRWWB) in \cite{Bacharach2019Hybrid}, which combines continuous-parameter scores with discrete-state tests.

The real channel vector \(\mathbf g(\boldsymbol\vartheta,\mathbf z)\) stacks the real and imaginary parts of \(\vecop(\mathbf H_G)\), and \(\widehat{\mathbf g}\) is formed from \(\widehat{\mathbf H}_G\) in the same way. Applying the covariance inequality in \cite[eq.~(4)]{Bacharach2019Hybrid} to the channel error gives
\begin{equation}
\mathbb E_{\mathbf y_G,\mathbf z}[(\widehat{\mathbf g}-\mathbf g)
(\widehat{\mathbf g}-\mathbf g)^T]
\succeq\mathbf B_{\rm CRWW}=\mathbf C\mathbf V^\dagger\mathbf C^T,
\label{eq:hybrid_general_bound}
\end{equation}
where \(\mathbf C=\mathbb E_{\mathbf y_G,\mathbf z}[(\widehat{\mathbf g}-\mathbf g)\mathbf v^T]\) and \(\mathbf V=\mathbb E_{\mathbf y_G,\mathbf z}[\mathbf v\mathbf v^T]\). We construct the test vector as \(\mathbf v=[\mathbf v_{\boldsymbol\vartheta}^T,\mathbf v_Z^T]^T\), with \(\mathbf v_{\boldsymbol\vartheta}\) corresponding to the continuous parameters and \(\mathbf v_Z\) to the discrete visibility states.

For Gaussian receiver noise with variance \(\lambda_n\), the joint density is \(f(\mathbf y_G,\mathbf z;\boldsymbol\vartheta)=p_{\rm M}(\mathbf z)\,\mathcal{CN}(\mathbf y_G;\boldsymbol\mu,\lambda_n\mathbf I_{G_rP})\), with \(\boldsymbol\mu=\vecop[\mathbf H_G(\boldsymbol\vartheta,\mathbf z)\mathbf S]\). For \(\mathbf v_{\boldsymbol\vartheta}\), we differentiate the log joint density with respect to \(\boldsymbol\vartheta\). For \(\mathbf v_Z\), we use the Weiss--Weinstein construction in \cite{Bacharach2019Hybrid} with exponent \(1/2\), comparing the joint densities before and after a visibility bit is flipped. This gives
\begin{align}
\mathbf v_{\boldsymbol\vartheta}
&=\frac{\partial\log f}
{\partial\boldsymbol\vartheta}
=\frac{2}{\lambda_n}\operatorname{Re}
\left\{\left(
\frac{\partial\boldsymbol\mu}{\partial\boldsymbol\vartheta^T}
\right)^H(\mathbf y_G-\boldsymbol\mu)\right\},
\notag\\[-0.2em]
v_{Z,j}
&=(1-2z_j)\sqrt{
\frac{f(\mathbf y_G,T_j\mathbf z;\boldsymbol\vartheta)}
{f(\mathbf y_G,\mathbf z;\boldsymbol\vartheta)}},
\label{eq:hybrid_visibility_test}
\end{align}
where \(T_j\mathbf z\) changes only \(z_j\) to \(1-z_j\), and the tests are defined for flip pairs with positive prior probabilities.

Substituting the test vectors in \eqref{eq:hybrid_visibility_test} into the definitions of \(\mathbf C\) and \(\mathbf V\), we approximate \(\mathbb E_{\mathbf z}\) using the sample average over \(N_z\) states drawn from \(p_{\rm M}(\mathbf z)\), instead of enumerating all \(2^{\dim(\mathbf z)}\) states. For \(\mathbb E_{\mathbf y_G,\mathbf z}\), we evaluate the Gaussian integrals over \(\mathbf y_G\) analytically for each sampled \(\mathbf z\) and then average the results. Assuming that the channel estimator is locally unbiased with respect to \(\boldsymbol\vartheta\) after averaging over visibility and noise, and that differentiation and integration can be interchanged, we obtain
\begin{align}
\mathbf C
&=\left[\mathbb E_{\mathbf z}\!\left[
\frac{\partial\mathbf g}{\partial\boldsymbol\vartheta^T}\right],\,
-\mathbb E_{\mathbf y_G,\mathbf z}[\mathbf g\mathbf v_Z^T]\right],
\notag\\[-0.2em]
\mathbf V
&=\begin{bmatrix}
\mathbb E_{\mathbf y_G,\mathbf z}[\mathbf v_{\boldsymbol\vartheta}\mathbf v_{\boldsymbol\vartheta}^T]&
\mathbb E_{\mathbf y_G,\mathbf z}[\mathbf v_{\boldsymbol\vartheta}\mathbf v_Z^T]\\
\mathbb E_{\mathbf y_G,\mathbf z}[\mathbf v_Z\mathbf v_{\boldsymbol\vartheta}^T]&
\mathbb E_{\mathbf y_G,\mathbf z}[\mathbf v_Z\mathbf v_Z^T]
\end{bmatrix}.
\label{eq:hybrid_channel_matrices}
\end{align}
The two blocks of \(\mathbf C\) describe channel changes with the continuous parameters and visibility states, respectively. These expressions depend only on the joint model. The off-diagonal blocks of \(\mathbf V\) retain the coupling between the two types of unknowns.

Substituting \eqref{eq:hybrid_channel_matrices} into \eqref{eq:hybrid_general_bound} gives the channel-error matrix bound. Since \(\|\widehat{\mathbf g}-\mathbf g\|_2^2=\|\widehat{\mathbf H}_G-\mathbf H_G\|_F^2\), taking its trace and normalizing by the mean channel energy yields
\begin{equation}
\mathrm{HCRWWB}_G
=\frac{\operatorname{tr}(\mathbf C\mathbf V^\dagger\mathbf C^T)}
{\mathbb E_{\mathbf z}\|\mathbf H_G(\boldsymbol\vartheta,\mathbf z)\|_F^2},
\label{eq:block_bcrb}
\end{equation}
which is the channel-NMSE lower bound for the above estimator class under the reference model and serves as the performance reference in Section~V. The derivation of \eqref{eq:hybrid_general_bound} and the closed-form expressions for the blocks of \(\mathbf C\) and \(\mathbf V\) are given in Appendix~\ref{app:hcrwwb_derivation}.

\subsection{Computational Complexity}

We analyze the main steps of Algorithm~\ref{alg:proposed_sbl} with the array responses given. Let \(M_o=G_rP\) and \(D_0=G_tQ_L+Q_rQ_t\) be the observation and full coefficient dimensions. Denote the largest numbers of candidate LOS rows, receive atoms, transmit atoms, and NLOS coefficients over the outer rounds by \(K_L\), \(K_r\), \(K_t\), and \(K_N\), respectively. The candidate coefficient dimension is at most \(D_C=G_tK_L+K_N\).

Steps~\ref{algline:sbl} and~\ref{algline:hyper} estimate the coefficients by SBL. Using the Woodbury identity, each iteration solves an \(M_o\times M_o\) system and updates the posterior statistics, with complexity \(\mathcal O(M_o^3+M_o^2D)\). The initial run uses \(D=D_0\), while each later refit uses \(D\le D_C\). With \(I_0\) initial iterations and at most \(I_C\) iterations per refit, the SBL cost over at most \(T_z\) refits is
\begin{equation}
\begin{aligned}
\mathcal C_{\rm SBL}=\mathcal O\bigl(&I_0(M_o^3+M_o^2D_0)\\
&+T_zI_C(M_o^3+M_o^2D_C)\bigr).
\end{aligned}
\label{eq:sbl_complexity}
\end{equation}

Steps~\ref{algline:receive} and~\ref{algline:los} update receive and LOS visibility by Viterbi recursion. For each atom, forming the group costs takes \(\mathcal O(M_o)\), and recursion and traceback take \(\mathcal O(G_r)\). Step~\ref{algline:transmit} tests each transmit state using all \(M_o\) observations. Including residual and prediction updates, one receive sweep costs \(\mathcal O(K_rM_o+K_NG_r)\), one transmit sweep costs \(\mathcal O(K_tG_tM_o+K_NG_tP)\), and one LOS sweep costs \(\mathcal O(K_LM_o)\).

Each pass through steps~\ref{algline:receive}--\ref{algline:los} contains at most \(I_r\) receive sweeps, followed by one transmit sweep and one LOS sweep. With at most \(I_v\) passes per visibility-update round and at most \(T_z\) such rounds, the visibility cost is
\begin{equation}
\begin{aligned}
\mathcal C_{\rm vis}=\mathcal O\bigl(T_zI_v[&
I_r(K_rM_o+K_NG_r)\\
&+(K_tG_t+K_L)M_o+K_NG_tP]\bigr).
\end{aligned}
\label{eq:visibility_total_complexity}
\end{equation}
The SBL iterations stop at tolerance \(\epsilon\), while the visibility sweeps stop when the states no longer change. Their numbers depend on the initialization, SNR, candidate support, and Markov transition probabilities.

Adding the initial transmit pilot products, the prediction and residual setup for each visibility round, and the final channel reconstruction in step~\ref{algline:reconstruct} gives
\begin{equation}
\begin{aligned}
&\mathcal C_{\rm total}=\mathcal C_{\rm SBL}+\mathcal C_{\rm vis}
+\mathcal O\bigl(Q_tG_tP\\[-0.2em]
&\quad{}+T_z(D_CM_o+K_tG_tP)+G_r[K_N+G_t(K_t+K_L)]\bigr).
\end{aligned}
\label{eq:algorithm_total_complexity}
\end{equation}
Grouping reduces the observation dimension \(M_o\), while candidate screening reduces the coefficient dimension from \(D_0\) to at most \(D_C\) in later refits. Both reduce the cost of the SBL updates.

\FloatBarrier
\section{Numerical Results}
% One top figure per column keeps each page to at most two numbered figures.
\setcounter{topnumber}{1}
In this section, we evaluate wavenumber-domain recovery and channel-estimation NMSE under different SNRs, pilot numbers, transmitter--receiver distances, and antenna-group counts.

\subsection{Simulation Setup}

We consider a \(50\)-GHz system with extended scattering clusters and limited blockage. The system has \(N_t=96\) transmit antennas and \(N_r=72\) receive antennas, with half-wavelength spacing (approximately \(3\) mm). For rectangular UPAs with at least two elements along each dimension, changing the aspect ratio at these fixed antenna counts gives transmit and receive aperture diagonals of \(3.91\)--\(14.10\) cm and \(3.19\)--\(10.50\) cm, respectively, measured between element centers. The corresponding conventional Rayleigh distances range from \(0.51\) to \(6.63\) m and from \(0.34\) to \(3.68\) m, respectively \cite{Balanis2016}.

Unless otherwise stated, \((G_t,G_r)=(8,24)\). The distance between the transmitter and receiver is \(R_0=1\) m. The channel includes one LOS path and three equal-power extended NLOS scattering clusters, each modeled as a disk of radius \(0.03\) m parallel to the UPA planes. The total NLOS-to-LOS power ratio is \(-3\) dB. Their nominal centers are \((-0.030,-0.020,0.28)\), \((0.020,0.025,0.50)\), and \((0.050,-0.030,0.72)\) m. Each disk center has an independent zero-mean Gaussian position offset with standard deviation \(0.002\) m per coordinate. The receive and transmit wavenumber grids contain \(Q_r=48\) and \(Q_t=24\) points, respectively. The resulting NLOS coefficient matrix is divided into \(3\times4\) blocks.

The transmitter adopts DFT-based pilots, and the rows of the pilot matrix are mutually orthogonal when \(P\ge G_t\). We consider insufficient and sufficient pilots, where the sufficient-pilot case satisfies \(P>G_t\). The received SNR is defined as \(\mathrm{SNR}=10\log_{10}\!\bigl(\|\mathbf H_G\mathbf S\|_F^2/(G_rP\lambda_n)\bigr)\) in decibels, where \(\lambda_n\) is the receiver-noise variance. The NMSE is
\begin{equation}
\mathrm{NMSE}=
\frac{\sum_{i=1}^{N_{\rm mc}}\|\widehat{\mathbf H}_{G,i}-\mathbf H_{G,i}\|_F^2}
{\sum_{i=1}^{N_{\rm mc}}\|\mathbf H_{G,i}\|_F^2},
\label{eq:nmse_def}
\end{equation}
where \(\mathbf H_{G,i}\) and \(\widehat{\mathbf H}_{G,i}\) are the true and estimated grouped channels in trial \(i\), and \(N_{\rm mc}=100\) is the number of trials.

All estimators use the grouped observations in \eqref{eq:Y_receive_grouped}. We compare LS, spatial near-field (NF) OMP based on the geometric NLOS dictionary in \cite{DaiMixed2023}, wavenumber block-OMP, and the proposed method with and without visibility estimation. For spatial NF OMP, coarse angle and distance ranges covering the scattering clusters are provided to construct the spatial dictionary. LS estimates the grouped channel without a sparse prior, using the minimum-norm solution when pilots are insufficient. Wavenumber block-OMP selects blocks from the wavenumber dictionary. For the proposed method without visibility estimation, \(\mathbf Z_L\), \(\mathbf Z_r\), and \(\mathbf Z_t\) are fixed to all-one matrices, so all retained atoms are assumed visible to all antenna groups. The HCRWWB in \eqref{eq:block_bcrb} is plotted as a performance reference.
% The present SNR on/off implementations also differ in their terminal coefficient refits; do not describe their difference as a visibility-only ablation.
The Markov priors in \eqref{eq:markov_prior_main}--\eqref{eq:markov_stationary_initial} use \(T^u_{0,1}=T^u_{1,0}=0.3\), \(u\in\{L,r,t\}\).

\subsection{Wavenumber-Domain Recovery and Channel-Estimation Performance}

\subsubsection{Wavenumber-Domain Block Sparsity}

\begin{figure}[t]
\centering
\includegraphics[width=0.98\linewidth]{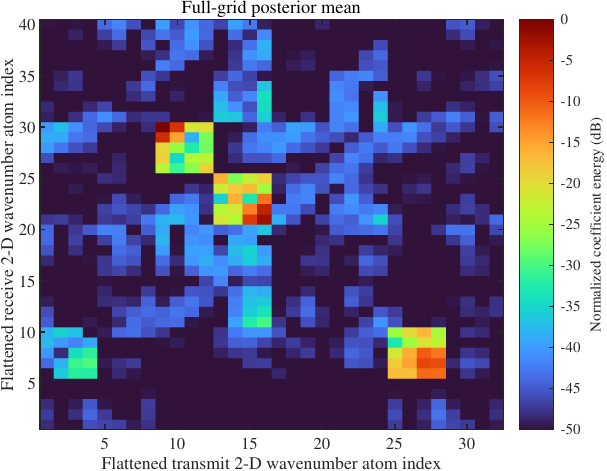}
\caption{Estimated NLOS block-sparse coefficients in the wavenumber domain.}
\label{fig:block_sparsity_estimated}
\end{figure}

\begin{figure}[t]
\centering
\includegraphics[width=0.98\linewidth]{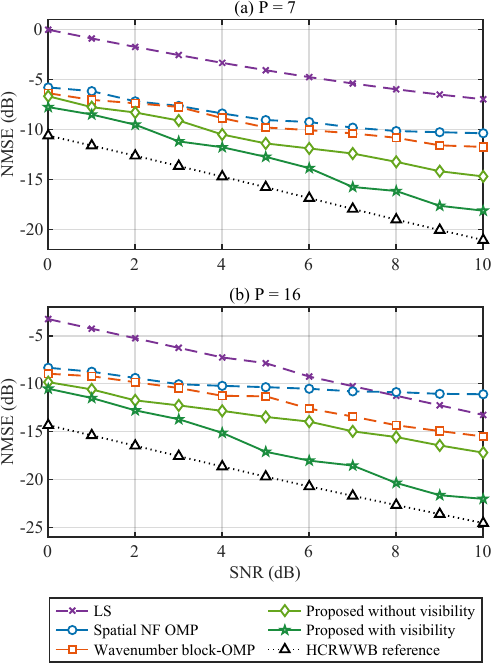}
\caption{NMSE versus SNR.}
\label{fig:snr_r8}
\end{figure}

Fig.~\ref{fig:block_sparsity_estimated} illustrates the localized block structure of the recovered NLOS coefficients. We use \(P=16\) and SNR \(=15\) dB.
Most of the recovered energy lies in a few local regions, consistent with Proposition~\ref{prop:finite_aperture_localization_main}. An extended cluster occupies a limited range of propagation directions and hence a limited receive--transmit wavenumber region. The direction varies across the apertures and within the cluster, so its energy spreads over neighboring atoms rather than a single grid point. This structure motivates the shared block variances in the proposed estimator: neighboring coefficients describe the same cluster and should be estimated together.

\begin{figure}[t]
\centering
\includegraphics[width=0.98\linewidth]{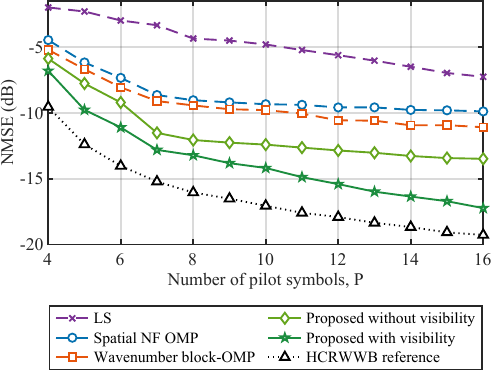}
\caption{NMSE versus the number of pilot symbols.}
\label{fig:pilot_nmse}
\end{figure}

\begin{figure}[t]
\centering
\includegraphics[width=0.98\linewidth]{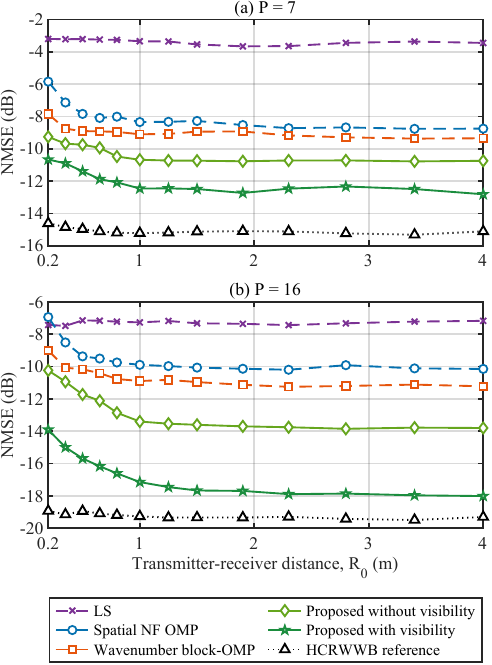}
\caption{NMSE versus the transmitter--receiver distance.}
\label{fig:distance_nmse}
\end{figure}

\subsubsection{Effect of SNR}

Fig.~\ref{fig:snr_r8} compares the NMSE with insufficient pilots (\(P=7\)) and sufficient pilots (\(P=16\)) at \(R_0=1\) m.

(i) The NMSEs of all methods decrease as SNR increases, owing to the reduction in the noise variance \(\lambda_n\). With sufficient pilots, the transmit-group contributions can be separated, so lower noise improves coefficient and visibility estimation more effectively. With insufficient pilots, the pilot matrix cannot fully separate these contributions, which limits the improvement from reducing noise.

(ii) Spatial NF OMP, which combines the spatial-domain model in \cite{DaiMixed2023} with OMP, performs worse than wavenumber block-OMP under both pilot settings. Its geometric dictionary is constructed from candidate point-scatterer responses, whereas the simulated NLOS paths are generated by disk-shaped extended clusters with spatially varying visibility. This difference causes model mismatch, and the Taylor approximation in the spatial-domain model introduces additional approximation error. Although coarse angle and distance ranges covering the clusters are provided, a finite spatial grid still cannot fully represent their distributed responses. Refining this grid increases the dictionary size, atom correlation, and computational cost, whereas a coarse grid increases off-grid error. In contrast, the proposed wavenumber model represents each extended cluster by a localized coefficient block without estimating every scatterer location separately, which better matches the simulated channel.

(iii) The advantage of the proposed method without visibility estimation can be explained from two aspects. First, compared with wavenumber block-OMP, block-SBL repeatedly updates the coefficients and shared block variances, allowing incorrect blocks to be suppressed. In contrast, early support errors in greedy block selection can affect subsequent recovery. Second, the proposed method jointly estimates \(\mathbf W_L\) and \(\mathbf X_N\), whereas the method in \cite{DaiMixed2023} estimates LOS and NLOS in two stages. With a total NLOS-to-LOS power ratio of \(-3\) dB, NLOS is not negligible during the first-stage LOS estimation. Joint estimation reduces the propagation of LOS estimation errors into NLOS recovery.

(iv) Visibility estimation further reduces the NMSE. Without visibility estimation, \(\mathbf Z_L\), \(\mathbf Z_r\), and \(\mathbf Z_t\) are fixed to all-one matrices, so every retained path is assumed visible to all antenna groups. The visibility-aware method removes the predicted contributions in blocked groups and uses the Markov prior to exploit the correlation between neighboring visibility states. It therefore describes the spatially varying channel more accurately than coefficient sparsity alone.

\subsubsection{Effect of the Number of Pilot Symbols}

Fig.~\ref{fig:pilot_nmse} demonstrates the effect of the pilot number at SNR \(=5\) dB. The visibility-aware method has the lowest NMSE among the estimators over the tested range. The structured methods improve more rapidly with insufficient pilots and more slowly once the pilots become sufficient, i.e., \(P>G_t=8\).

With insufficient pilots, additional independent pilots reveal previously unobserved transmit combinations. They reduce the ambiguity between candidate channel responses and make both coefficient and visibility estimation more reliable. At \(P=G_t\), the pilot matrix becomes full row rank. When \(P>G_t\), further pilots will not provide new observation dimension, which explains the reduced slopes. LS benefits from the added observations but does not exploit cluster structure, whereas the spatial NF OMP curve flattens earlier because more pilots cannot remove dictionary mismatch.

The proposed method uses the added observations to refine distributed block coefficients instead of making only greedy support decisions. Its visibility-aware variant can also distinguish weak visible responses from blocked groups more reliably. These two effects explain why additional pilots remain useful even after the transmit-group contributions can be separated.

\subsubsection{Effect of Transmitter--Receiver Distance}

Fig.~\ref{fig:distance_nmse} examines the effect of the transmitter--receiver distance at SNR \(=5\) dB for \(P=7\) and \(P=16\), with \(R_0\) ranging from \(0.2\) to \(4.0\) m. The structured estimators show an initial NMSE decrease followed by a much smaller change, whereas LS varies little. Since the received SNR is fixed, this trend is not a gain from increased received power.

For fixed aperture and cluster sizes, increasing the distance narrows the wavenumber regions in \eqref{eq:Krq_main} and \eqref{eq:Ktq_main}. The dominant energy is therefore concentrated in fewer neighboring wavenumber coefficients, which facilitates block-sparse channel recovery. For spatial NF OMP, increasing the distance reduces the approximation error of the finite geometric model.

At larger distances, propagation gradually approaches far-field behavior, and the wavenumber support varies only slightly with distance. The NMSE curves therefore become nearly flat. The proposed wavenumber-domain model and estimator remain applicable in both the near- and far-field regions, and the visibility-aware method achieves the lowest NMSE over the tested distance range.

\begin{figure}[t]
\centering
\includegraphics[width=0.98\linewidth]{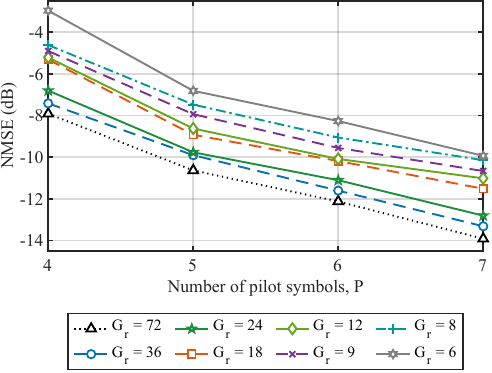}
\caption{NMSE versus the number of pilot symbols for different receive-group counts.}
\label{fig:grouping_effect}
\end{figure}

\begin{figure}[t]
\centering
\includegraphics[width=0.98\linewidth]{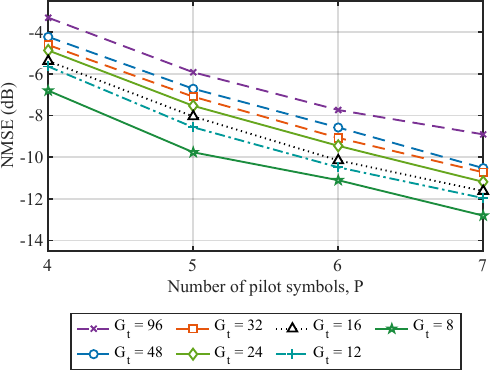}
\caption{NMSE versus the number of pilot symbols for different transmit-group counts.}
\label{fig:transmit_grouping_effect}
\end{figure}

\subsection{Effect of Antenna Grouping}

Figs.~\ref{fig:grouping_effect} and~\ref{fig:transmit_grouping_effect} exhibit the accuracy--complexity tradeoff caused by receive/transmit grouping. In Fig.~\ref{fig:grouping_effect}, we fix \(G_t=8\), SNR \(=5\) dB, and \(P\in\{4,5,6,7\}\).
Finer receive grouping achieves a lower NMSE because each group contains fewer antennas, reducing the variation represented by a single group-center response and a shared visibility state. It also retains more receive observations for separating cluster contributions. The elementwise case \(G_r=72\) yields the lowest NMSE, but increases the observation dimension and the SBL cost in Section~IV-D. With \(G_r=24\), the NMSE remains within about \(1\) dB of the elementwise reference over the tested pilot range while using one third as many receive groups. We therefore adopt \(G_r=24\) in the other experiments.

Fig.~\ref{fig:transmit_grouping_effect} shows the opposite effect at the transmitter: finer grouping increases the NMSE under the tested insufficient-pilot settings. Increasing \(G_t\) introduces more LOS coefficients in \(\mathbf W_L\) and more transmit visibility states, but does not add observations when \(G_r\) and \(P\) are fixed. More transmit contributions must therefore be distinguished from the same pilot measurements. Here, this loss outweighs the gain from a more accurate group-center approximation.
The curve for \(G_t=8\) also decreases faster with \(P\). As \(P\) approaches \(8\), the pilots observe a larger fraction of the transmit dimensions; for much larger \(G_t\), they remain insufficient over the entire tested range. Thus, \(G_t=8\) gives the best accuracy in this comparison and keeps the coefficient and visibility dimensions small.

\setcounter{topnumber}{2}
\section{Conclusion}

This paper developed an electromagnetic wavenumber-domain model for massive MIMO channels with scattering clusters and spatial visibility. By accounting for scattering clusters and partial visibility over large arrays, the model describes more realistic propagation scenarios and applies to both near- and far-field regions. Based on the local block structure of the cluster responses, we proposed a visibility-aware block-SBL estimator that alternates coefficient estimation with Markov visibility updates. An HCRWWB was also derived as an NMSE reference for the continuous channel parameters and discrete visibility states.

The simulation results show that the visibility-aware method achieves lower NMSE than LS, spatial NF OMP, and wavenumber block-OMP. Increasing SNR improves estimation, while additional pilots give larger gains when pilots are insufficient than after they become sufficient. The distance results further show that the proposed model and estimator apply to both near- and far-field channels. Finer receive grouping improves accuracy, whereas finer transmit grouping can worsen estimation when the pilot number is fixed and insufficient. These results support \((G_t,G_r)=(8,24)\) as an accuracy--complexity compromise for the considered system.

\appendices
\section{Derivation and Evaluation of the HCRWWB}
\label{app:hcrwwb_derivation}

For the reference model in \eqref{eq:hybrid_complete_observation}, we first establish a general channel-error bound, then construct the continuous and discrete test vectors, and finally evaluate \(\mathbf C\) and \(\mathbf V\). All transmit and receive antenna positions are known in a common coordinate system. The continuous unknowns are the common LOS gain and the cluster gains and positions. For differentiation, each complex gain is represented by its real and imaginary parts, giving \(\dim(\boldsymbol\vartheta)=2+5Q_c\). The propagation distances and directions are determined by the antenna positions and \(\mathbf c_q\), rather than varied independently.

The physical visibility vector contains \(G_rG_t\) LOS group-pair states, \(G_rQ_c\) receive-cluster states, and \(G_tQ_c\) transmit-cluster states. Thus, \(\dim(\mathbf z)=G_rG_t+Q_c(G_r+G_t)\). The Markov prior is independent of \(\boldsymbol\vartheta\), with positive initial and transition probabilities, so \(p_{\rm M}(\mathbf z)>0\) for every binary state. The deterministic parameter vector \(\boldsymbol\vartheta\) is fixed during state averaging. The given receiver-noise variance \(\lambda_n>0\) is also held fixed during differentiation, state flips, and state averaging; it is not recalculated for each flipped state. The receiver noise satisfies \(\vecop(\mathbf N_{\rm rx})\sim\mathcal{CN}(\mathbf0,\lambda_n\mathbf I_{N_rP})\), independently of \(\mathbf z\). Since the receive groups are disjoint and \(\mathbf R_r\mathbf R_r^H=\mathbf I_{G_r}\), \(\vecop(\mathbf R_r\mathbf N_{\rm rx})\) has covariance \(\lambda_n\mathbf I_{G_rP}\).

\subsection{General Channel-Error Bound}

The channel error is \(\mathbf e=\widehat{\mathbf g}(\mathbf y_G)-\mathbf g(\boldsymbol\vartheta,\mathbf z)\), with error matrix \(\mathbf R_e=\mathbb E_{\mathbf y_G,\mathbf z}[\mathbf e\mathbf e^T]\). At this stage, \(\mathbf v\) is an arbitrary real test vector. Assume that \(\mathbf e\) and \(\mathbf v\) have finite second moments, and form \(\mathbf C=\mathbb E_{\mathbf y_G,\mathbf z}[\mathbf e\mathbf v^T]\) and \(\mathbf V=\mathbb E_{\mathbf y_G,\mathbf z}[\mathbf v\mathbf v^T]\). Following the covariance argument in \cite[eq.~(4)]{Bacharach2019Hybrid}, for any real matrix \(\mathbf A\) of compatible dimensions,
\begin{equation}
\begin{aligned}
\mathbf0
&\preceq\mathbb E_{\mathbf y_G,\mathbf z}
  [(\mathbf e-\mathbf A\mathbf v)(\mathbf e-\mathbf A\mathbf v)^T]\\
&=\mathbf R_e-\mathbf A\mathbf C^T-\mathbf C\mathbf A^T
  +\mathbf A\mathbf V\mathbf A^T.
\end{aligned}
\label{eq:app_hcr_covariance_expansion}
\end{equation}
This also covers a singular \(\mathbf V\). Indeed, \(\mathbf V\mathbf a=\mathbf0\) implies \(\mathbb E_{\mathbf y_G,\mathbf z}[(\mathbf a^T\mathbf v)^2]=0\), hence \(\mathbf a^T\mathbf v=0\) almost surely and \(\mathbf C\mathbf a=\mathbf0\). Consequently, \(\mathbf C=\mathbf C\mathbf V^\dagger\mathbf V\). Setting \(\mathbf A=\mathbf C\mathbf V^\dagger\) in \eqref{eq:app_hcr_covariance_expansion} gives
\begin{equation}
\begin{aligned}
&\mathbf R_e-\mathbf C\mathbf V^\dagger\mathbf C^T\\
&\quad=\mathbb E_{\mathbf y_G,\mathbf z}
\big[(\mathbf e-\mathbf C\mathbf V^\dagger\mathbf v)
     (\mathbf e-\mathbf C\mathbf V^\dagger\mathbf v)^T\big]
\succeq\mathbf0.
\end{aligned}
\label{eq:app_hcr_projection_bound}
\end{equation}
Equation~\eqref{eq:app_hcr_projection_bound} establishes the general channel-error bound in \eqref{eq:hybrid_general_bound}. However, \(\mathbf C\) can still depend on the estimator through \(\mathbf e\). We therefore construct test vectors that remove this dependence under the local unbiasedness condition stated below.

\subsection{Construction of the Test Vectors}

Following \cite{Bacharach2019Hybrid}, we combine the continuous-parameter score with discrete-state tests as \(\mathbf v=[\mathbf v_{\boldsymbol\vartheta}^T,\mathbf v_Z^T]^T\). For compactness, write the joint density as \(f_{\mathbf z}=f(\mathbf y_G,\mathbf z;\boldsymbol\vartheta)\). Since the prior does not depend on \(\boldsymbol\vartheta\), differentiating the Gaussian log likelihood gives
\begin{equation}
\begin{aligned}
\mathbf v_{\boldsymbol\vartheta}
&=\frac{\partial\log f_{\mathbf z}}{\partial\boldsymbol\vartheta}
=-\frac{1}{\lambda_n}
\frac{\partial\|\mathbf y_G-\boldsymbol\mu\|_2^2}
{\partial\boldsymbol\vartheta}\\
&=\frac{2}{\lambda_n}\operatorname{Re}\!\left\{
\left(\frac{\partial\boldsymbol\mu}{\partial\boldsymbol\vartheta^T}\right)^H
(\mathbf y_G-\boldsymbol\mu)\right\}.
\end{aligned}
\label{eq:app_hcr_continuous_test}
\end{equation}

For the discrete part, we specialize the Weiss--Weinstein test in \cite{Bacharach2019Hybrid} to exponent \(1/2\) and a unit displacement \(\mathbf e_j\) in the \(j\)-th binary coordinate. With the joint density set to zero outside the binary state space, the two-sided test becomes
\begin{equation}
\begin{aligned}
v_{Z,j}
&=\sqrt{\frac{f_{\mathbf z+\mathbf e_j}}{f_{\mathbf z}}}
-\sqrt{\frac{f_{\mathbf z-\mathbf e_j}}{f_{\mathbf z}}}\\
&=(1-2z_j)\sqrt{\frac{f_{T_j\mathbf z}}{f_{\mathbf z}}}.
\end{aligned}
\label{eq:app_hcr_discrete_test}
\end{equation}
Only the forward term remains when \(z_j=0\), whereas only the negative backward term remains when \(z_j=1\). This gives the flip test in \eqref{eq:hybrid_visibility_test}. Both test vectors depend on the observations and the statistical model, but neither involves \(\widehat{\mathbf g}\).

\subsection{Evaluation of \(\mathbf C\) and \(\mathbf V\)}

For fixed \(\mathbf z\), the conditional Gaussian likelihood is \(p_{\mathbf z}(\mathbf y_G)=f_{\mathbf z}/p_{\rm M}(\mathbf z)\). The joint expectation of any integrable matrix- or scalar-valued function \(F\) satisfies
\begin{equation}
\begin{aligned}
&\mathbb E_{\mathbf y_G,\mathbf z}[F(\mathbf y_G,\mathbf z)]\\
&\quad=\sum_{\mathbf z}p_{\rm M}(\mathbf z)
\int F(\mathbf y_G,\mathbf z)p_{\mathbf z}(\mathbf y_G)\,d\mathbf y_G.
\end{aligned}
\label{eq:app_hcr_joint_average}
\end{equation}
We first integrate over the continuous observation \(\mathbf y_G\) for each fixed \(\mathbf z\), and then sum over the binary states with weights \(p_{\rm M}(\mathbf z)\). This procedure applies to both the continuous and discrete test vectors.

With these test vectors, partition \(\mathbf C=[\mathbf C_{\boldsymbol\vartheta},\mathbf C_Z]\) and
\begin{equation}
\mathbf V=
\begin{bmatrix}
\mathbf V_{\boldsymbol\vartheta\boldsymbol\vartheta}
&\mathbf V_{\boldsymbol\vartheta Z}\\
\mathbf V_{\boldsymbol\vartheta Z}^T&\mathbf V_{ZZ}
\end{bmatrix}.
\label{eq:app_hcr_V_partition}
\end{equation}
Here, \(\mathbf C\) has \(2G_rG_t\) rows and \(\dim(\boldsymbol\vartheta)+\dim(\mathbf z)\) columns, and \(\mathbf V\) is square with the latter dimension.

\subsubsection{Continuous Block of \(\mathbf C\)}
The averaged bias is
\begin{equation}
\mathbf b(\boldsymbol\vartheta)
=\mathbb E_{\mathbf y_G,\mathbf z}
 [\widehat{\mathbf g}-\mathbf g(\boldsymbol\vartheta,\mathbf z)].
\label{eq:app_hcr_bias}
\end{equation}
Local unbiasedness at the parameter point under consideration requires \(\mathbf b=\mathbf0\) and \(\partial\mathbf b/\partial\boldsymbol\vartheta^T=\mathbf0\). The estimator \(\widehat{\mathbf g}(\mathbf y_G)\) does not explicitly depend on the unknown \(\boldsymbol\vartheta\). Assuming a differentiable channel response and interchangeability of differentiation and integration, we obtain
\begin{equation}
\begin{aligned}
\frac{\partial\mathbf b}{\partial\boldsymbol\vartheta^T}
&=\sum_{\mathbf z}\int\left[
\mathbf e\frac{\partial f}{\partial\boldsymbol\vartheta^T}
-\frac{\partial\mathbf g}{\partial\boldsymbol\vartheta^T}f
\right]d\mathbf y_G\\
&=\mathbf C_{\boldsymbol\vartheta}
-\mathbb E_{\mathbf z}\!\left[
\frac{\partial\mathbf g}{\partial\boldsymbol\vartheta^T}\right],
\end{aligned}
\label{eq:app_hcr_bias_derivative}
\end{equation}
where \(\partial f/\partial\boldsymbol\vartheta^T=f\mathbf v_{\boldsymbol\vartheta}^T\). With \(\partial\mathbf b/\partial\boldsymbol\vartheta^T=\mathbf0\), the observation integral and state average give
\begin{equation}
\begin{aligned}
\mathbf C_{\boldsymbol\vartheta}
&=\sum_{\mathbf z}p_{\rm M}(\mathbf z)
\frac{\partial\mathbf g}{\partial\boldsymbol\vartheta^T}
\int p_{\mathbf z}(\mathbf y_G)\,d\mathbf y_G\\
&=\sum_{\mathbf z}p_{\rm M}(\mathbf z)
\frac{\partial\mathbf g}{\partial\boldsymbol\vartheta^T}
=\mathbb E_{\mathbf z}\!\left[
\frac{\partial\mathbf g}{\partial\boldsymbol\vartheta^T}\right].
\end{aligned}
\label{eq:app_hcr_C_continuous}
\end{equation}
Here, the channel derivative is fixed during the observation integral, and \(\int p_{\mathbf z}(\mathbf y_G)\,d\mathbf y_G=1\). Local unbiasedness is imposed after averaging over visibility and noise, rather than separately for every visibility state.

\subsubsection{Discrete Block of \(\mathbf C\)}
Pairing \(\mathbf z\) with \(T_j\mathbf z\) in \eqref{eq:app_hcr_discrete_test} gives
\begin{equation}
\begin{aligned}
&f_{\mathbf z}v_{Z,j}(\mathbf y_G,\mathbf z)
+f_{T_j\mathbf z}v_{Z,j}(\mathbf y_G,T_j\mathbf z)\\
&\quad=(1-2z_j)\sqrt{f_{\mathbf z}f_{T_j\mathbf z}}
-(1-2z_j)\sqrt{f_{\mathbf z}f_{T_j\mathbf z}}=0.
\end{aligned}
\label{eq:app_hcr_pair_cancellation}
\end{equation}
Summing over the pairs at each fixed \(\mathbf y_G\) gives \(\mathbb E_{\mathbf z\mid\mathbf y_G}[v_{Z,j}]=0\). Hence,
\begin{equation}
\begin{aligned}
\mathbb E_{\mathbf y_G,\mathbf z}[\widehat{\mathbf g}v_{Z,j}]
&=\mathbb E_{\mathbf y_G}\!\left[
\widehat{\mathbf g}\,
\mathbb E_{\mathbf z\mid\mathbf y_G}[v_{Z,j}]\right]=\mathbf0,\\
[\mathbf C_Z]_{:,j}
&=-\mathbb E_{\mathbf y_G,\mathbf z}[\mathbf g v_{Z,j}].
\end{aligned}
\label{eq:app_hcr_C_discrete_identity}
\end{equation}
This cancellation does not require a discrete-state unbiasedness condition. Both blocks of \(\mathbf C\) are now independent of the particular estimator within the stated class.

For compactness, the fixed \(\boldsymbol\vartheta\) is omitted from \(\boldsymbol\mu(\mathbf z)\) and \(\mathbf g(\mathbf z)\) below. Define
\begin{equation}
\begin{aligned}
\sigma_j(\mathbf z)&=1-2z_j,&
\rho_j(\mathbf z)&=\frac{p_{\rm M}(T_j\mathbf z)}{p_{\rm M}(\mathbf z)},\\
\Delta\boldsymbol\mu_j(\mathbf z)
&=\boldsymbol\mu(T_j\mathbf z)-\boldsymbol\mu(\mathbf z),\\
\Delta\mathbf g_j(\mathbf z)
&=\mathbf g(T_j\mathbf z)-\mathbf g(\mathbf z).
\end{aligned}
\label{eq:app_hcr_flip_quantities}
\end{equation}
For the Markov prior, the ratio \(\rho_j\) involves only the transition factors adjacent to the flipped state and, if it is the first state of a sequence, its initial probability. All other factors cancel.

The Gaussian integrals follow from a common identity. For two states \(\mathbf z\) and \(\mathbf z'\), write their conditional means as \(\boldsymbol\mu\) and \(\boldsymbol\mu'\). Completing the square gives
\begin{equation}
\begin{aligned}
&\|\mathbf y_G-\boldsymbol\mu\|_2^2
+\|\mathbf y_G-\boldsymbol\mu'\|_2^2\\
&\quad=2\left\|\mathbf y_G-
\frac{\boldsymbol\mu+\boldsymbol\mu'}{2}\right\|_2^2
+\frac12\|\boldsymbol\mu'-\boldsymbol\mu\|_2^2,
\end{aligned}
\label{eq:app_hcr_complete_square}
\end{equation}
and thus
\begin{equation}
\begin{aligned}
\sqrt{p_{\mathbf z}(\mathbf y_G)p_{\mathbf z'}(\mathbf y_G)}
={}&\exp\!\left(-\frac{\|\boldsymbol\mu'-\boldsymbol\mu\|_2^2}
{4\lambda_n}\right)\\
&\times\mathcal{CN}\!\left(\mathbf y_G;
\frac{\boldsymbol\mu+\boldsymbol\mu'}{2},
\lambda_n\mathbf I_{G_rP}\right).
\end{aligned}
\label{eq:app_hcr_gaussian_product}
\end{equation}
The remaining Gaussian density integrates to one. In particular, the overlap of the original and flipped likelihoods is
\begin{equation}
B_j(\mathbf z)
=\int\sqrt{p_{\mathbf z}p_{T_j\mathbf z}}\,d\mathbf y_G
=\exp\!\left(-\frac{\|\Delta\boldsymbol\mu_j\|_2^2}{4\lambda_n}\right).
\label{eq:app_hcr_single_overlap}
\end{equation}

Pairing the states in \eqref{eq:app_hcr_C_discrete_identity}, factoring out their prior probabilities, and applying \eqref{eq:app_hcr_single_overlap} yields
\begin{equation}
\begin{aligned}
[\mathbf C_Z]_{:,j}
&=\sum_{\mathbf z:z_j=0}\Delta\mathbf g_j
\int\sqrt{f_{\mathbf z}f_{T_j\mathbf z}}\,d\mathbf y_G\\
&=\sum_{\mathbf z:z_j=0}
\sqrt{p_{\rm M}(\mathbf z)p_{\rm M}(T_j\mathbf z)}
B_j\Delta\mathbf g_j\\
&=\frac12\mathbb E_{\mathbf z}
\!\left[\sigma_j\sqrt{\rho_j}\,B_j\Delta\mathbf g_j\right].
\end{aligned}
\label{eq:app_hcr_C_discrete_closed}
\end{equation}
The sums restricted to \(z_j=0\) include each flip pair once, whereas the final expectation includes both members and therefore requires the factor \(1/2\). Equations~\eqref{eq:app_hcr_C_continuous} and \eqref{eq:app_hcr_C_discrete_closed} give the two blocks of \(\mathbf C\) from channel derivatives and visibility-induced channel differences, respectively.

\subsubsection{Continuous Block of \(\mathbf V\)}
For fixed \(\mathbf z\), the residual \(\mathbf y_G-\boldsymbol\mu\) is proper complex Gaussian with covariance \(\lambda_n\mathbf I\). Its real and imaginary parts each have covariance \((\lambda_n/2)\mathbf I\). Substituting \eqref{eq:app_hcr_continuous_test} and integrating this conditional second moment gives
\begin{equation}
\begin{aligned}
&\int\mathbf v_{\boldsymbol\vartheta}
\mathbf v_{\boldsymbol\vartheta}^T
p_{\mathbf z}(\mathbf y_G)\,d\mathbf y_G\\
&\quad=\frac{2}{\lambda_n}\operatorname{Re}\!\left\{
\left(\frac{\partial\boldsymbol\mu}{\partial\boldsymbol\vartheta^T}\right)^H
\frac{\partial\boldsymbol\mu}{\partial\boldsymbol\vartheta^T}
\right\}.
\end{aligned}
\label{eq:app_hcr_continuous_conditional_moment}
\end{equation}
Averaging over the visibility states then yields
\begin{equation}
\begin{aligned}
\mathbf V_{\boldsymbol\vartheta\boldsymbol\vartheta}
&=\sum_{\mathbf z}p_{\rm M}(\mathbf z)
\int\mathbf v_{\boldsymbol\vartheta}\mathbf v_{\boldsymbol\vartheta}^T
p_{\mathbf z}(\mathbf y_G)\,d\mathbf y_G\\
&=\frac{2}{\lambda_n}\mathbb E_{\mathbf z}
\!\left[\operatorname{Re}\!\left\{
\left(\frac{\partial\boldsymbol\mu}{\partial\boldsymbol\vartheta^T}\right)^H
\frac{\partial\boldsymbol\mu}{\partial\boldsymbol\vartheta^T}
\right\}\right].
\end{aligned}
\label{eq:app_hcr_V_continuous}
\end{equation}
The observation derivative follows directly from the channel model:
\begin{equation}
\frac{\partial\boldsymbol\mu}{\partial\boldsymbol\vartheta^T}
=(\mathbf S^T\otimes\mathbf I_{G_r})
\frac{\partial\vecop(\mathbf H_G)}{\partial\boldsymbol\vartheta^T}.
\label{eq:app_hcr_channel_derivative}
\end{equation}
Stacking the real and imaginary parts of the channel derivative gives \(\partial\mathbf g/\partial\boldsymbol\vartheta^T\) in \eqref{eq:app_hcr_C_continuous}.

\subsubsection{Continuous--Discrete Block of \(\mathbf V\)}
For each fixed \(\mathbf z\), \eqref{eq:app_hcr_gaussian_product} reduces the required observation integral to the first moment of a Gaussian density with midpoint mean:
\begin{equation}
\int[\mathbf y_G-\boldsymbol\mu(\mathbf z)]
\sqrt{p_{\mathbf z}p_{T_j\mathbf z}}\,d\mathbf y_G
=\frac12B_j\Delta\boldsymbol\mu_j.
\label{eq:app_hcr_midpoint_moment}
\end{equation}
Substituting the two tests, evaluating this integral, and then averaging over \(\mathbf z\) gives
\begin{equation}
\begin{aligned}
[\mathbf V_{\boldsymbol\vartheta Z}]_{:,j}
&=\sum_{\mathbf z}p_{\rm M}(\mathbf z)\sigma_j\sqrt{\rho_j}
\int\mathbf v_{\boldsymbol\vartheta}
\sqrt{p_{\mathbf z}p_{T_j\mathbf z}}\,d\mathbf y_G\\
&=\frac{1}{\lambda_n}\sum_{\mathbf z}p_{\rm M}(\mathbf z)
\sigma_j\sqrt{\rho_j}\,B_j\\
&\qquad\times\operatorname{Re}\!\left\{
\left(\frac{\partial\boldsymbol\mu}{\partial\boldsymbol\vartheta^T}\right)^H
\Delta\boldsymbol\mu_j\right\}.
\end{aligned}
\label{eq:app_hcr_V_cross}
\end{equation}
The derivative is evaluated at the original state \(\mathbf z\). The factor \(1/2\) in \eqref{eq:app_hcr_midpoint_moment} cancels the factor \(2\) in the continuous score.

\subsubsection{Discrete Block of \(\mathbf V\)}
Multiplying the two discrete tests cancels their common denominator in the joint expectation, so that
\begin{equation}
\begin{aligned}
[\mathbf V_{ZZ}]_{ij}
&=\sum_{\mathbf z}p_{\rm M}(\mathbf z)
\sigma_i\sigma_j\sqrt{\rho_i\rho_j}\\
&\qquad\times\int\sqrt{p_{T_i\mathbf z}p_{T_j\mathbf z}}\,d\mathbf y_G\\
&=\mathbb E_{\mathbf z}
\!\left[\sigma_i\sigma_j\sqrt{\rho_i\rho_j}\,B_{ij}\right],
\end{aligned}
\label{eq:app_hcr_V_discrete}
\end{equation}
where the overlap between the two flipped likelihoods is
\begin{equation}
B_{ij}(\mathbf z)
=\exp\!\left(-\frac{\|\Delta\boldsymbol\mu_i
-\Delta\boldsymbol\mu_j\|_2^2}{4\lambda_n}\right).
\label{eq:app_hcr_double_overlap}
\end{equation}
For \(i=j\), \(B_{jj}=1\) and
\begin{equation}
[\mathbf V_{ZZ}]_{jj}
=\mathbb E_{\mathbf z}[\rho_j]
=\sum_{\mathbf z}p_{\rm M}(T_j\mathbf z)=1.
\label{eq:app_hcr_discrete_diagonal}
\end{equation}
Both the continuous--discrete block and the off-diagonal entries of \(\mathbf V_{ZZ}\) are retained.

\subsection{Evaluation for the Continuous Green Model}

The derivatives and state differences above are evaluated from the continuous Green terms in \eqref{eq:Enlos_aperture_continuous} and \eqref{eq:Elos_aperture_continuous}, as specified in the reference model. The visibility states are shared within each group. In particular,
\begin{equation}
\begin{aligned}
\nu_{u,q}^A(\boldsymbol\rho_{u,n_u})
&=\nu_{u,q}^A(\boldsymbol\rho_{u,g_u(n_u)}),\quad u\in\{r,t\},\\
\nu_L^A(\boldsymbol\rho_{r,n_r},\boldsymbol\rho_{t,n_t})
&=\nu_L^A(\boldsymbol\rho_{r,g_r(n_r)},
\boldsymbol\rho_{t,g_t(n_t)}).
\end{aligned}
\label{eq:app_hcr_physical_visibility_map}
\end{equation}
Only the visibility is shared: the Green functions are evaluated at the individual antenna positions. The group membership, pilot matrix, and port-conversion matrix remain fixed.

For fixed \(\mathbf z\), differentiation with respect to the gain components gives
\begin{align}
\frac{\partial E_{\mathcal A}^{\rm LOS}}
{\partial\operatorname{Re}\alpha_L}
&=-\nu_L^A(\boldsymbol\rho_r,\boldsymbol\rho_t)
\frac{\jj\omega_c\mu_b}{4\pi}\notag\\[-0.2em]
&\quad\cdot\frac{e^{-\jj k_c\|\mathbf r_r(\boldsymbol\rho_r)-\mathbf r_t(\boldsymbol\rho_t)\|_2}}
{\|\mathbf r_r(\boldsymbol\rho_r)-\mathbf r_t(\boldsymbol\rho_t)\|_2},
\label{eq:app_hcr_los_gain_derivative}\\
\frac{\partial E_{\mathcal A}^{\rm NLOS}}
{\partial\operatorname{Re}\alpha_q}
&=-\nu_{r,q}^A(\boldsymbol\rho_r)\nu_{t,q}^A(\boldsymbol\rho_t)
\chi_{r,q}(\boldsymbol\rho_r)\chi_{t,q}(\boldsymbol\rho_t).
\label{eq:app_hcr_nlos_gain_derivative}
\end{align}
For either field, the derivative with respect to the imaginary part of its gain is \(\jj\) times the corresponding derivative with respect to the real part. Since the array positions are known, the LOS propagation distance is not an additional unknown.

Each cluster position affects both propagation segments. For \(u\in\{r,t\}\) and positive propagation distances, differentiating \eqref{eq:chi_propagation_factors} gives
\begin{equation}
\begin{aligned}
\nabla_{\mathbf c_q}\chi_{u,q}(\boldsymbol\rho_u)
={}&\chi_{u,q}(\boldsymbol\rho_u)
\left(\jj k_c+\frac{1}{\|\mathbf r_u(\boldsymbol\rho_u)-\mathbf c_q\|_2}\right)\\
&\cdot\frac{\mathbf r_u(\boldsymbol\rho_u)-\mathbf c_q}
{\|\mathbf r_u(\boldsymbol\rho_u)-\mathbf c_q\|_2}.
\end{aligned}
\label{eq:app_hcr_green_position_derivative}
\end{equation}
The NLOS field derivative therefore follows from the product rule:
\begin{equation}
\begin{aligned}
\nabla_{\mathbf c_q}E_{\mathcal A}^{\rm NLOS}
={}&-\alpha_q\nu_{r,q}^A(\boldsymbol\rho_r)\nu_{t,q}^A(\boldsymbol\rho_t)\\
&\cdot\bigl[\chi_{t,q}(\boldsymbol\rho_t)
\nabla_{\mathbf c_q}\chi_{r,q}(\boldsymbol\rho_r)\\
&\qquad+\chi_{r,q}(\boldsymbol\rho_r)
\nabla_{\mathbf c_q}\chi_{t,q}(\boldsymbol\rho_t)\bigr].
\end{aligned}
\label{eq:app_hcr_field_position_derivative}
\end{equation}
Its three entries are the derivatives with respect to \(x_q\), \(y_q\), and \(z_q\). The visibility states are held fixed in these derivatives and are handled separately by the discrete tests.

For each real parameter \(\vartheta_a\), the grouped channel derivative is
\begin{equation}
\frac{\partial\mathbf H_G}{\partial\vartheta_a}
=\mathbf R_r\mathbf C_{\rm port}
\left[\frac{\partial(E_{\mathcal A}^{\rm LOS}+E_{\mathcal A}^{\rm NLOS})}
{\partial\vartheta_a}\right]_{n_r,n_t}\mathbf B_t,
\label{eq:app_hcr_physical_channel_derivative}
\end{equation}
where each \((n_r,n_t)\) entry is evaluated at the aperture coordinates satisfying \(\mathbf r_r(\boldsymbol\rho_{r,n_r})=\mathbf r_{n_r}\) and \(\mathbf r_t(\boldsymbol\rho_{t,n_t})=\mathbf r_{n_t}\).
Vectorizing these derivatives and applying \eqref{eq:app_hcr_channel_derivative} gives the observation derivative. Stacking their real and imaginary parts gives the channel derivative used in \(\mathbf C_{\boldsymbol\vartheta}\). Both blocks therefore use the same physical parameters.

For a discrete test, \(T_j\mathbf z\) flips one physical group-visibility state in \eqref{eq:app_hcr_physical_visibility_map}. A receive- or transmit-cluster flip changes that cluster's response for every antenna in the selected group. An LOS flip changes the direct response for the selected receive--transmit group pair. Re-evaluating the same Green model gives
\begin{equation}
\begin{aligned}
\Delta\mathbf H_{G,j}
&=\mathbf H_G(\boldsymbol\vartheta,T_j\mathbf z)
-\mathbf H_G(\boldsymbol\vartheta,\mathbf z),\\
\Delta\boldsymbol\mu_j&=\vecop(\Delta\mathbf H_{G,j}\mathbf S),\\
\Delta\mathbf g_j
&=\begin{bmatrix}
\operatorname{Re}\vecop(\Delta\mathbf H_{G,j})\\
\operatorname{Im}\vecop(\Delta\mathbf H_{G,j})
\end{bmatrix}.
\end{aligned}
\label{eq:app_hcr_physical_flip_differences}
\end{equation}
Substituting these differences and the above derivatives into the five matrix-block expressions completes their evaluation for the reference model. No finite-wavenumber atom or independently varied segment distance is used in this calculation.

\subsection{Channel NMSE Bound and Numerical Evaluation}

Equations~\eqref{eq:app_hcr_C_continuous}, \eqref{eq:app_hcr_C_discrete_closed}, \eqref{eq:app_hcr_V_continuous}, \eqref{eq:app_hcr_V_cross}, and \eqref{eq:app_hcr_V_discrete} specify all blocks of \(\mathbf C\) and \(\mathbf V\). Substitution into \eqref{eq:app_hcr_projection_bound}, followed by taking the trace, gives
\begin{equation}
\begin{aligned}
\mathbb E_{\mathbf y_G,\mathbf z}
 [\|\widehat{\mathbf H}_G-\mathbf H_G\|_F^2]
&=\operatorname{tr}(\mathbf R_e)\\
&\ge\operatorname{tr}(\mathbf C\mathbf V^\dagger\mathbf C^T).
\end{aligned}
\label{eq:app_hcr_channel_trace_bound}
\end{equation}
Normalizing by \(\mathbb E_{\mathbf z}\|\mathbf H_G\|_F^2>0\) gives \eqref{eq:block_bcrb}. The error, the derivatives, the flipped channels, and the normalizing energy all refer to the same Green reference channel. The field-modeling residuals are not included in this bound or absorbed into \(\lambda_n\).

All observation integrals have now been evaluated analytically. For any resulting matrix- or scalar-valued integrand \(F(\mathbf z)\), the remaining discrete average is approximated using \(N_z\) independent states \(\mathbf z^{(n)}\sim p_{\rm M}\):
\begin{equation}
\begin{aligned}
\mathbb E_{\mathbf z}[F(\mathbf z)]
&=\sum_{\mathbf z}p_{\rm M}(\mathbf z)F(\mathbf z)\\
&\approx\frac{1}{N_z}\sum_{n=1}^{N_z}F(\mathbf z^{(n)}).
\end{aligned}
\label{eq:app_hcr_state_average}
\end{equation}
Because the states are sampled from \(p_{\rm M}\), each sampled contribution has weight \(1/N_z\). This average is applied to the five matrix blocks above, retaining the factor \(1/2\) in \eqref{eq:app_hcr_C_discrete_closed}. The mean channel energy is computed in the same way:
\begin{equation}
\begin{aligned}
\mathbb E_{\mathbf z}\|\mathbf H_G(\boldsymbol\vartheta,\mathbf z)\|_F^2
\approx\frac{1}{N_z}\sum_{n=1}^{N_z}
\|\mathbf H_G(\boldsymbol\vartheta,\mathbf z^{(n)})\|_F^2.
\end{aligned}
\label{eq:app_hcr_sampled_channel_energy}
\end{equation}
The matrix blocks and channel energy are averaged before computing the pseudoinverse and the ratio in \eqref{eq:block_bcrb}; the per-state bounds are not averaged. This avoids enumerating all binary states and provides a numerical approximation to the bound defined by the exact expectations.

\end{document}